\documentclass[aps,prx,reprint,showpacs,amsmath,amssymb,superscriptaddress,floatfix,longbibliography]{revtex4-2}

\usepackage{graphicx}
\usepackage[export]{adjustbox}
\usepackage{dcolumn}
\usepackage{upgreek}
\usepackage{bm,dsfont}
\usepackage{amssymb,mathtools}
\usepackage{booktabs}
\usepackage{color}
\usepackage{microtype}
\usepackage[T1]{fontenc}
\usepackage{lmodern}
\usepackage[shortlabels]{enumitem}
\usepackage{physics}
\usepackage{pgfplots}
\usepackage{tikz}
\usepackage{placeins}
\usepackage[colorlinks,plainpages=false,pdfpagemode=UseNone,pdfstartview=FitBH]{hyperref}

\pgfplotsset{compat=1.18}

\allowdisplaybreaks
\hypersetup{
colorlinks = true,
linkcolor = [rgb]{0.87,0.18,0.22},
citecolor = [rgb]{0.0,0.60,0.32},
urlcolor = [rgb]{0.20, 0.30, 0.54}}

\newcommand{\vecbf}[1]{\ensuremath{\boldsymbol{#1}}}

\renewcommand{\cp}{{\mathrm{i}}}

\newcommand{\mytrace}[1]{ \operatorname{Tr}\bigl[#1\bigr] }

\newcommand{\wexp}[1]{\langle #1 \rangle_{\!\Omega}}
\newcommand{\gaugePKernel}{\ensuremath{\hat{\Lambda}(\vecbf{x})}}

\DeclareMathOperator{\diag}{diag}

\begin{document}

\title{Neural Gauge-\texorpdfstring{$P$}{P} Representation for Open Quantum Dynamics of Interacting Bosons}

\author{Xiaodong Cao}
\email{xdcao@ustc.edu.cn}
\affiliation{Suzhou Institute for Advanced Research, University of Science and Technology of China, Suzhou 215123, China}
\affiliation{School of Artificial Intelligence and Data Science, University of Science and Technology of China, Suzhou 215123, China}

\author{Zhicheng Zhong}
\email{zczhong@ustc.edu.cn}
\affiliation{Suzhou Institute for Advanced Research, University of Science and Technology of China, Suzhou 215123, China}
\affiliation{School of Artificial Intelligence and Data Science, University of Science and Technology of China, Suzhou 215123, China}

\date{\today}

\begin{abstract}
    Simulating the nonequilibrium dynamics of interacting open quantum systems
    remains challenging beyond small system sizes.  Quantum phase-space
    representations provide a scalable approach, but their useful simulation
    time can be limited by broad distribution tails and the associated
    boundary terms.  We introduce the neural gauge-$P$ representation for open
    bosonic systems, in which stochastic gauges are parameterized by neural
    networks and optimized using exact moment equation residuals.  For the
    driven-dissipative Bose--Hubbard model in both single-site and
    square-lattice settings, the neural gauge-$P$ representation remains
    accurate during long-time evolution toward the steady state, whereas the
    corresponding ungauged representation becomes unreliable at substantially
    earlier times.  These results demonstrate the potential of the neural
    gauge-$P$ representation for accurate simulations of
    nonequilibrium open quantum many-body dynamics.
\end{abstract}
\maketitle

%======================================================================
\section{Introduction}\label{sec:intro}
%======================================================================
Interacting open quantum many-body systems combine coherent evolution,
external driving, dissipation, and quantum fluctuations.  Their competition
can produce phenomena without an equilibrium counterpart, including
dissipative phase transitions~\cite{Minganti_spectral_theory_Liouvillians_2018,Sieberer_Keldysh_open_quantum_systems_2016,Carusotto_quantum_fluids_of_light_2013},
critical dynamics and metastability~\cite{Sieberer_dynamical_critical_phenomena_2013,Macieszczak_metastability_open_quantum_dynamics_2016},
and reservoir-engineered many-body states~\cite{Diehl_open_cold_atoms_2008,verstraete_states_engineering_2009}.
Open bosonic systems provide natural settings for this physics.  Examples
include cavity QED~\cite{Raimond_manipulating_2001,Walther_cavity_quantum_electrodynamics_2006,Reiserer_cavity_network_2015},
circuit QED~\cite{Schmidt_circuit_QED_lattices_2013,Carusotto_circuit_QED_2020},
and exciton-polariton lattices~\cite{Hopfield_excitons_1958,Kim_polariton_condensation_2011,Klembt_polariton_TI_2018,Whittaker_polariton_Lieb_lattice_2018,Goblot_polariton_flatband_2019,Su_polariton_condensation_2020}.
Driven-dissipative bosonic devices also support applications in quantum
information, such as stabilized cat states and Kerr-cat qubits~\cite{Mirrahimi_protected_cat_qubits_2014,Leghtas_two_photon_loss_2015,Grimm_Kerr_cat_qubit_2020,Gautier_cat_qubit_2022,Chamberland_concatenated_cat_codes_2022,Pan_phase_space_compression_2023,Hajr_Kerr_cat_qubit_2D_architecture_2024,Hutin_neural_network_cat_states_2025,Yuan_weak_Kerr_nonlinearities_2025,Ding_Kerr_cat_qubit_2025}.
These developments require reliable methods for real-time dynamics and
nonequilibrium steady states in interacting open bosonic systems.

Several numerical approaches have made promising progress in this direction.  Tensor-network methods provide controlled descriptions of many one-dimensional and quasi-two-dimensional open systems~\cite{Verstraete_MPS_finite_temperature_dissipative_2004,Cui_MPS_steady_state_2015,Werner_positive_tensor_network_approach_2016,Kilda_projected_entangled_pair_operator_2021,Hryniuk_variational_Monte_Carlo_approach_2024,Godinez_Riemannian_approach_2025,Sander_large_scale_stochastic_simulation_2025}. For bosons, however, the unbounded local Hilbert space and the fast growth entanglement make long-time dynamics difficult. Neural density-matrix methods offer expressive ans\"atze for open-system density operators, using the flexibility of neural networks to represent complex entanglement and correlation structures~\cite{Torlai_neural_density_operators_2018,Hartmann_neural_network_approach_2019,Vicentini_variational_neural_network_ansatz_2019,Nagy_vmc_neural_network_ansatz_2019,Kothe_liouville_space_neural_network_representation_2024,Mellak_neural_networks_variational_solutions_2024,Lin_neural_density_operators_2024}.  Most demonstrations, however, have focused on spin models or other systems
with finite-dimensional local Hilbert spaces, while applications to large
open bosonic systems remain limited, despite recent progress for closed-system ground states~\cite{Denis_accurate_neural_quantum_states_2025}.  Corner-space renormalization (CSR) method constructs reduced low-rank representations of the many-body density matrix and can reach high accuracy for steady states, but they are primarily designed for steady-state properties and are most efficient when the relevant density matrix has moderately low von Neumann entropy~\cite{Finazzi_corner_space_2015,Rota_quadratic_driven_2D_lattice_2019}.

Phase-space methods take a complementary approach and avoid an explicit
local Fock-space cutoff.  The positive-$P$ representation expands the
density matrix in a doubled coherent-state phase space and maps the Lindblad
equation to stochastic differential equations (SDEs).  The number of
phase-space variables grows only linearly with system size~\cite{Drummond_positive_P_1980,Deuar_positive_P_2006,Deuar_positive_P_2021}.
This mapping relies on integration by parts and is exact only when the
associated boundary terms vanish.  Nonlinear dynamics can generate broad
phase-space tails that violate this condition.  The same broad tails produce rare,
large trajectory contributions, which can dominate finite-sample averages
and cause spiking estimates.

To mitigate these limitations, gauge-$P$ representations exploit the
overcompleteness of coherent states to introduce drift and diffusion
gauges, generating an infinite family of gauge-equivalent kernel equations
with different stochastic dynamics and sampling
properties~\cite{Plimak_diffusion_gauge_2001,Deuar_gauge_P_2002,
Deuar_gauge_P_2006}.  Analytic gauges can partially delay or suppress
spiking trajectories in simple models and restricted parameter
regimes~\cite{Plimak_diffusion_gauge_2001,Drummond_positive_P_1980,
Deuar_positive_P_2006,Wuster_positive_P_and_gauge_P_2017}, but their
construction is usually tailored to a particular system and does not
readily generalize.  Moreover, drift gauges can reduce large phase-space
excursions at the cost of enhanced fluctuations in the stochastic weight.
Balancing these competing effects is
therefore challenging and requiring flexible parameterization of the stochastic gauges for general systems.

In this work, we introduce the neural gauge-$P$ representation, in which
the drift and diffusion gauge functions are parameterized by neural
networks.  We optimize these stochastic gauges by minimizing residuals of the exact
equations of motion for normally ordered moments.  These exact moment equation residuals quantify the corresponding projected
boundary terms and provide necessary conditions for consistency with the
exact Lindblad dynamics. The paper is organized as follows.  In
Sec.~\ref{sec:model_and_method:bh}, we introduce the driven-dissipative
Bose--Hubbard model.  Section~\ref{sec:model_and_method:gauge_p} develops
the neural gauge-$P$ representation and its stochastic equations, while
Sec.~\ref{sec:model_and_method:neural_gauge_optimization} establishes the
connection between projected boundary terms and exact moment equation
residuals and constructs the corresponding optimization objective.  Details of the graph neural
network (GNN) used in this work are given in
Appendix~\ref{app:neural_architecture}.  In Sec.~\ref{sec:results}, we first study the undriven and coherently
driven single-site cases, comparing the results with exact diagonalization,
analyzing the learned drift and diffusion gauges, and examining sample-size convergence.  We then apply the method to a large square lattice and
benchmark the long-time results against corner-space-renormalization reference. We conclude and outline future directions in Sec.~\ref{sec:conclusion}.

%======================================================================
\section{Model and method}\label{sec:model_and_method}
%======================================================================

\begin{figure*}[!t]
    \centering
    \includegraphics[width=\textwidth]{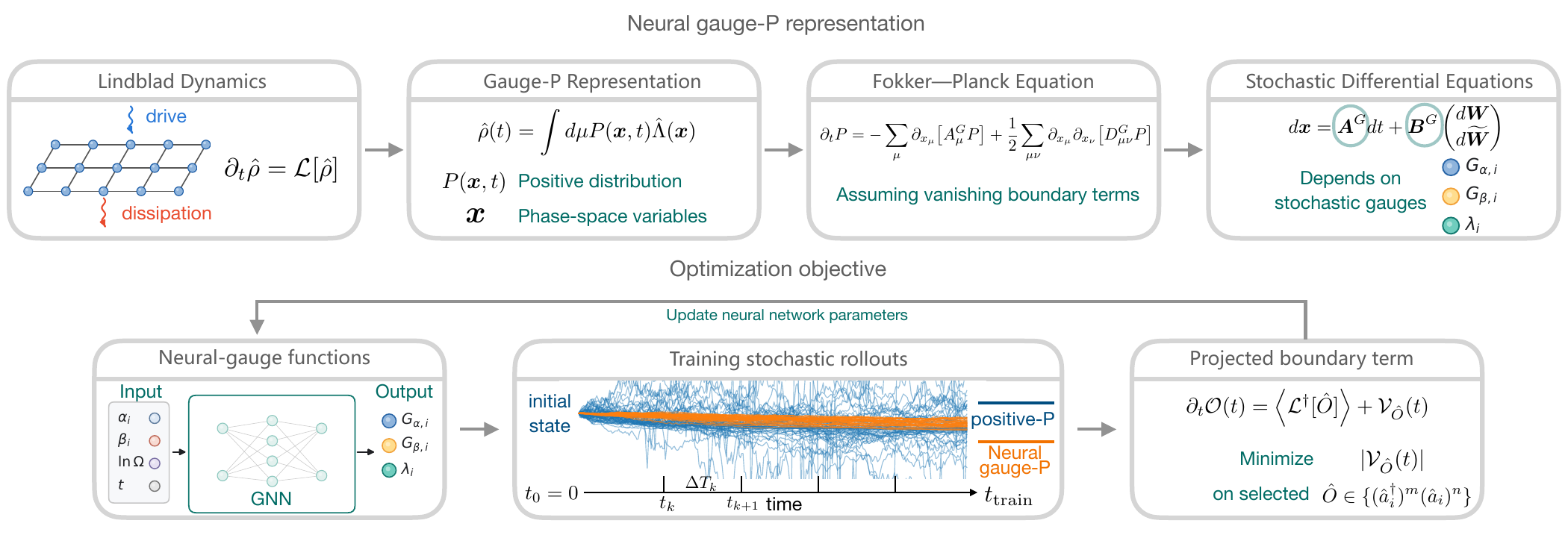}
    \caption{
        Neural gauge-$P$ representation.  The gauge-$P$ expansion maps the
        Lindblad equation to a Fokker--Planck equation when the boundary terms
        generated by integration by parts vanish.  The corresponding SDEs
        contain drift and diffusion gauges.  A graph neural network maps the
        phase-space variables $\vecbf\alpha$ and $\vecbf\beta$, the complex log
        weight $\ln\Omega$, and time $t$ to the local drift gauges
        $G_{\alpha_i}$ and $G_{\beta_i}$ and diffusion gauge $\lambda_i$.
        Training rollouts are scored by covariance-normalized residuals of
        exact moment equations, which probe selected projections of the
        boundary term.  The residual loss is then used to update the network parameters.
    }
    \label{fig:neural_gauge_schematic}
\end{figure*}
In this section, we formulate the neural gauge-$P$ representation for the
driven-dissipative Bose--Hubbard model.  The overall workflow is summarized
in Fig.~\ref{fig:neural_gauge_schematic}.  We first introduce the model, then derive its neural gauge-$P$
representation, and finally construct the optimization objective from
covariance-normalized exact moment equation residuals.

%----------------------------------------------------------------------
\subsection{Driven-dissipative Bose--Hubbard model}
\label{sec:model_and_method:bh}
%----------------------------------------------------------------------

We consider a lattice of $M$ interacting bosonic modes, with the annihilation and creation operators collected into the vectors
$\hat{\vecbf{a}}=(\hat{a}_1,\dots,\hat{a}_M)^T$
and
$\hat{\vecbf{a}}^\dagger=(\hat{a}_1^\dagger,\dots,\hat{a}_M^\dagger)$,
respectively.
They satisfy the canonical commutation relations
$[\hat{a}_i,\hat{a}_j^\dagger]=\delta_{ij}$.
For simplicity, we consider homogeneous parameters, although the formulation extends straightforwardly to site-dependent cases.
The Hamiltonian reads
\begin{align}\label{eq:hamiltonian}
\hat{H}
=
\sum_{i=1}^{M}\left[
-\Delta \hat{n}_i
+ \frac{U}{2}\hat{a}_i^{\dagger 2}\hat{a}_i^2
+ F\hat{a}_i^\dagger
+ F^*\hat{a}_i
\right]
-
\hat{\vecbf{a}}^\dagger
\vecbf{J}
\hat{\vecbf{a}},
\end{align}
where $\hat{n}_i=\hat{a}_i^\dagger\hat{a}_i$ is the local density operator, $\Delta$ is the pump-cavity detuning, $U$ is the strength of the onsite Kerr nonlinearity, and $F$ is the coherent drive amplitude.
The hopping matrix $\vecbf{J}\in\mathbb{C}^{M\times M}$ is Hermitian,
$\vecbf{J}^\dagger=\vecbf{J}$, and the term
$-\hat{\vecbf{a}}^\dagger\vecbf{J}\hat{\vecbf{a}}$
describes hopping between lattice sites.  In the lattice calculations below, we consider a square lattice with
periodic boundary conditions and uniform nearest-neighbor hopping, such that
$J_{ij}=J$ for nearest-neighbor sites $i$ and $j$, and $J_{ij}=0$ otherwise.

In the presence of weak coupling to a Markovian environment, the system density operator evolves according to the Lindblad master equation~\cite{Lindblad_generators_1976,Gorini_completely_positive_1976}.
For local one-body loss, it takes the form
\begin{align}\label{eq:lindblad}
\partial_t\hat{\rho}
=
\mathcal{L}[\hat{\rho}]
=
-\cp\bigl[\hat{H},\hat{\rho}\bigr]
+
\gamma\sum_{i=1}^{M}
\left[
\hat{a}_i\hat{\rho}\hat{a}_i^\dagger
-
\frac{1}{2}
\left\{
\hat{a}_i^\dagger\hat{a}_i,
\hat{\rho}
\right\}
\right],
\end{align}
where $\gamma$ is the local one-body loss rate and $\mathcal{L}$ denotes the Liouvillian superoperator.

%----------------------------------------------------------------------
\subsection{Neural gauge-\texorpdfstring{$P$}{P} representation}
\label{sec:model_and_method:gauge_p}
%----------------------------------------------------------------------

For the $M$-mode bosonic system, we employ the gauge-$P$ representation,
which augments the doubled phase space of the positive-$P$ representation
with a complex weight $\Omega$ and incorporates additional stochastic
gauge freedoms~\cite{Plimak_diffusion_gauge_2001,Deuar_gauge_P_2002,
Deuar_gauge_P_2006}.

We begin with the unnormalized Bargmann coherent states
\begin{align}\label{eq:coherent_state}
\ket{\vecbf{\alpha}}
=
\bigotimes_{i=1}^{M}
e^{\alpha_i\hat a_i^\dagger}\ket{0}.
\end{align}
These states form an overcomplete basis of the bosonic Hilbert space, with
overlap
$\braket{\vecbf{\beta}^*}{\vecbf{\alpha}}
=\prod_{i=1}^{M}e^{\beta_i\alpha_i}$.
In the doubled phase space, $\vecbf\alpha$ and $\vecbf\beta$ are
treated as independent complex variables.  Together with the complex weight $\Omega$,
we collect them into
$\vecbf{x}
=(\alpha_1,\beta_1,\ldots,\alpha_M,\beta_M,\Omega)^T
\in\mathbb C^{2M+1}$.
The gauge-$P$ kernel is then defined as
\begin{align}\label{eq:gauge_P_kernel}
\gaugePKernel
=
\Omega
\frac{
\ket{\vecbf{\alpha}}\bra{\vecbf{\beta}^*}
}{
\braket{\vecbf{\beta}^*}{\vecbf{\alpha}}
},
\end{align}
which is analytic in all phase-space variables. 
Then, the many-body density operator can be represented as
\begin{align}\label{eq:gauge_P_expansion}
\hat\rho(t)
=
\int d\mu\,
P(\vecbf{x},t)\,
\gaugePKernel,
\end{align}
where
$d\mu
=\left(\prod_{i=1}^{M}d^2\alpha_i\,d^2\beta_i\right)d^2\Omega$
is the integration measure, and $P(\vecbf{x},t)\geq0$ is a real,
normalized distribution on the enlarged phase space satisfying
$\int d\mu\,P(\vecbf{x},t)=1$.
Since $\mytrace{\gaugePKernel}=\Omega$, the trace of the represented
density operator is
$\mytrace{\hat\rho(t)}
=\int d\mu\,P(\vecbf{x},t)\Omega$.

The analyticity of the gauge-$P$ kernel gives the differential
identities~\cite{Drummond_positive_P_1980,Deuar_positive_P_2006,
Deuar_gauge_P_2002}
\begin{align}\label{eq:differential_identities}
\hat a_i\gaugePKernel
&=
\alpha_i\gaugePKernel,
\nonumber\\
\hat a_i^\dagger\gaugePKernel
&=
\left(
\partial_{\alpha_i}+\beta_i
\right)\gaugePKernel,
\nonumber\\
\gaugePKernel\hat a_i
&=
\left(
\partial_{\beta_i}+\alpha_i
\right)\gaugePKernel,
\nonumber\\
\gaugePKernel\hat a_i^\dagger
&=
\beta_i\gaugePKernel.
\end{align}
Substituting Eq.~\eqref{eq:gauge_P_expansion} into the Lindblad equation
and using these differential identities gives
\begin{align}\label{eq:lindblad_kernel}
\partial_t\hat\rho
&=
\int d\mu\,
P(\vecbf{x},t)\,
\mathcal L^P[\gaugePKernel],
\nonumber\\
\mathcal L^P[\gaugePKernel]
&=
\sum_{\mu=1}^{2M}
A_\mu^P(\vecbf{x})\,
\partial_{x_\mu}\gaugePKernel
\nonumber\\
&\quad+
\frac{1}{2}
\sum_{\mu,\nu=1}^{2M}
D_{\mu\nu}^P(\vecbf{x})\,
\partial_{x_\mu}\partial_{x_\nu}\gaugePKernel.
\end{align}
Here $\mu$ and $\nu$ run over the $2M$ coordinates
$(\alpha_1,\beta_1,\ldots,\alpha_M,\beta_M)$, and the differential
operator does not yet act on $\Omega$.
The positive-$P$ drift vector is
$\vecbf A^P=(A_{\alpha_1}^P,A_{\beta_1}^P,\ldots,
A_{\alpha_M}^P,A_{\beta_M}^P)^T$, with components
\begin{align}\label{eq:p_drift}
A_{\alpha_i}^P
&=
\cp\Delta\alpha_i
-\cp U\alpha_i^2\beta_i
-\cp F
-\frac{\gamma}{2}\alpha_i
+\cp(\vecbf J\vecbf\alpha)_i,
\nonumber\\
A_{\beta_i}^P
&=
-\cp\Delta\beta_i
+\cp U\beta_i^2\alpha_i
+\cp F^*
-\frac{\gamma}{2}\beta_i
-\cp(\vecbf J^T\vecbf\beta)_i.
\end{align}
The corresponding diffusion matrix is diagonal,
$\vecbf D^P=\diag(D_{\alpha_1}^P,D_{\beta_1}^P,\ldots,
D_{\alpha_M}^P,D_{\beta_M}^P)$, with
$D_{\alpha_i}^P=-\cp U\alpha_i^2$ and
$D_{\beta_i}^P=\cp U\beta_i^2$.

A convenient factorization of the diffusion matrix is
$\vecbf D^P=\vecbf B\vecbf B^T$, where
$\vecbf B=\diag(B_{\alpha_1},B_{\beta_1},\ldots,
B_{\alpha_M},B_{\beta_M})$,
$B_{\alpha_i}=\sqrt{-\cp U}\,\alpha_i$, and
$B_{\beta_i}=\sqrt{\cp U}\,\beta_i$. This factorization is not unique.  For real diffusion-gauge functions $\vecbf\lambda(\vecbf{x},t)=(\lambda_1(\vecbf{x},t),\ldots,\lambda_M(\vecbf{x},t))$, we introduce
the block-diagonal complex-orthogonal matrix
\begin{align}\label{eq:diffusion_gauge_rotation}
\vecbf R_{\lambda}
&=
\bigoplus_{i=1}^{M}\vecbf r_i,
\nonumber\\
\vecbf r_i
&=
\begin{pmatrix}
\cosh(\lambda_i) & \cp\sinh(\lambda_i)\\
-\cp\sinh(\lambda_i) & \cosh(\lambda_i)
\end{pmatrix}.
\end{align}
Since
$\vecbf R_{\lambda}\vecbf R_{\lambda}^{T}=\vecbf I_{2M}$,
the transformed noise-amplitude matrix
$\widetilde{\vecbf B}=\vecbf B\vecbf R_{\lambda}$ satisfies
$\widetilde{\vecbf B}\widetilde{\vecbf B}^{T}=\vecbf D^P$.
The diffusion gauge therefore redistributes the stochastic noise without
changing the diffusion matrix.

Although diffusion gauges can improve sampling efficiency, they do not
generally eliminate boundary terms~\cite{Plimak_diffusion_gauge_2001,
Deuar_gauge_P_2002,Deuar_gauge_P_2006}.
Further freedom is provided by drift gauges, which act through the
additional weight variable $\Omega$.
The gauge-$P$ kernel satisfies
\begin{align}\label{eq:omega_identity}
\left(
\Omega\partial_\Omega-1
\right)\gaugePKernel=0.
\end{align}
For arbitrary complex drift-gauge functions
$\vecbf G(\vecbf{x},t)
=(G_{\alpha_1}(\vecbf{x},t),G_{\beta_1}(\vecbf{x},t),\ldots,G_{\alpha_M}(\vecbf{x},t),G_{\beta_M}(\vecbf{x},t))^T$,
this identity implies
\begin{align}\label{eq:drift_gauge_identities}
\sum_{\nu=1}^{2M}
\left[
\frac{1}{2}G_\nu^2\Omega\partial_\Omega
+
G_\nu
\sum_{\mu=1}^{2M}
\widetilde B_{\mu\nu}\partial_{x_\mu}
\right]
\left(
\Omega\partial_\Omega-1
\right)\gaugePKernel
=0.
\end{align}
Adding Eq.~\eqref{eq:drift_gauge_identities} to the positive-$P$ kernel
equation, Eq.~\eqref{eq:lindblad_kernel}, gives the gauged
kernel equation
\begin{align}\label{eq:gauged_lindblad_kernel}
\partial_t\hat\rho
&=
\int d\mu\,
P(\vecbf{x},t)\,
\mathcal L^G[\gaugePKernel],
\nonumber\\
\mathcal L^G[\gaugePKernel]
&=
\sum_{\mu=1}^{2M+1}
A_\mu^G(\vecbf{x},t)\partial_{x_\mu}\gaugePKernel
\nonumber\\
&\quad+
\frac{1}{2}
\sum_{\mu,\nu=1}^{2M+1}
D_{\mu\nu}^G(\vecbf{x},t)
\partial_{x_\mu}\partial_{x_\nu}\gaugePKernel.
\end{align}
The gauged drift vector and noise-amplitude matrix are
\begin{align}\label{eq:full_gauge_p_drift}
\vecbf A^G
&=
\begin{pmatrix}
\vecbf A^P-\widetilde{\vecbf B}\vecbf G\\
0
\end{pmatrix},
\\
\vecbf B^G
&=
\begin{pmatrix}
\widetilde{\vecbf B}\\
\Omega\vecbf G^T
\end{pmatrix}.
\nonumber
\end{align}
The resulting diffusion matrix in the enlarged phase space is
\begin{align}\label{eq:full_gauge_p_diffusion}
\vecbf D^G
=
\vecbf B^G(\vecbf B^G)^T
=
\begin{pmatrix}
\vecbf D^P
&
\Omega\widetilde{\vecbf B}\vecbf G
\\
\Omega(\widetilde{\vecbf B}\vecbf G)^T
&
\Omega^2\vecbf G^T\vecbf G
\end{pmatrix}.
\end{align}

We next apply integration by parts to
Eq.~\eqref{eq:gauged_lindblad_kernel}.  Moving derivatives from the kernel
to $P(\vecbf{x},t)$ gives
\begin{align}\label{eq:explicit_integration_by_parts}
    \partial_t\hat\rho(t)
    &=
    \int d\mu\,
    P(\vecbf{x},t)\,
    \mathcal L^G[\gaugePKernel]
    \nonumber\\
    &=
    \int d\mu\,
    \gaugePKernel\,
    \mathcal L_{\rm FP}^G[P(\vecbf{x},t)]
    -
    \widehat{\mathcal V}(t),
    \end{align}
where the Fokker--Planck forward operator is
\begin{align}\label{eq:gauge_p_forward_operator}
\mathcal L_{\rm FP}^G[P(\vecbf{x},t)]
&=
-\sum_{\mu=1}^{2M+1}
\partial_{x_\mu}
\left[
A_\mu^G P(\vecbf{x},t)
\right]
\nonumber\\
&\quad+
\frac{1}{2}
\sum_{\mu,\nu=1}^{2M+1}
\partial_{x_\mu}\partial_{x_\nu}
\left[
D_{\mu\nu}^G P(\vecbf{x},t)
\right].
\end{align}
The corresponding operator-valued boundary term is
\begin{align}\label{eq:operator_boundary_term}
\widehat{\mathcal V}(t)
=-{}&
\int d\mu\,
\sum_{\mu=1}^{2M+1}
\partial_{x_\mu}
\Bigg[
A_\mu^G P(\vecbf{x},t)\gaugePKernel
\nonumber\\[-0.3ex]
&\quad
-\frac{1}{2}
\sum_{\nu=1}^{2M+1}
\partial_{x_\nu}
\left[
D_{\mu\nu}^G P(\vecbf{x},t)
\right]
\gaugePKernel
\nonumber\\[-0.3ex]
&\quad
+\frac{1}{2}
\sum_{\nu=1}^{2M+1}
D_{\mu\nu}^G P(\vecbf{x},t)
\partial_{x_\nu}\gaugePKernel
\Bigg].
\end{align}
Direct evaluation of Eq.~\eqref{eq:operator_boundary_term} requires the
high-dimensional phase-space distribution $P(\vecbf{x},t)$ and its
derivatives, and is therefore generally challenging.  Under the
vanishing-boundary-term assumption $\widehat{\mathcal V}(t)=0$,
Eq.~\eqref{eq:explicit_integration_by_parts} reduces to the gauge-$P$
Fokker--Planck equation
\begin{align}\label{eq:gauged_fpe}
\partial_tP(\vecbf{x},t)
=
\mathcal L_{\rm FP}^G
\left[
P(\vecbf{x},t)
\right].
\end{align}
Resolving the complex phase-space variables into their real and imaginary
parts yields a real positive-semidefinite diffusion matrix,
Eq.~\eqref{eq:gauged_fpe} therefore admits an equivalent It\^o SDE
realization~\cite{Deuar_gauge_P_2002}.

Let $d\vecbf W=(dW_1,\ldots,dW_M)^T$ and
$d\widetilde{\vecbf W}
=(d\widetilde W_1,\ldots,d\widetilde W_M)^T$ denote two independent sets
of real Wiener increments satisfying
\begin{align*}
\langle dW_i\,dW_j\rangle
&=
\langle d\widetilde W_i\,d\widetilde W_j\rangle
=
\delta_{ij}\,dt,
\\
\langle dW_i\,d\widetilde W_j\rangle
&=0.
\end{align*}
The corresponding gauge-$P$ SDEs are
\begin{align}\label{eq:gauge_P_sde}
d\alpha_i
&=
A_{\alpha_i}^G\,dt
+
B_{\alpha_i}\cosh(\lambda_i)\,dW_i
+
\cp B_{\alpha_i}\sinh(\lambda_i)\,d\widetilde W_i,
\nonumber\\
d\beta_i
&=
A_{\beta_i}^G\,dt
-
\cp B_{\beta_i}\sinh(\lambda_i)\,dW_i
+
B_{\beta_i}\cosh(\lambda_i)\,d\widetilde W_i,
\nonumber\\
d\Omega
&=
\Omega
\sum_{i=1}^{M}
\left(
G_{\alpha_i}\,dW_i
+
G_{\beta_i}\,d\widetilde W_i
\right),
\end{align}
where the gauged phase-space drifts are
\begin{align}\label{eq:gauge_P_sde_components}
A_{\alpha_i}^G
&=
A_{\alpha_i}^P
-
B_{\alpha_i}\cosh(\lambda_i)\,G_{\alpha_i}
-
\cp B_{\alpha_i}\sinh(\lambda_i)\,G_{\beta_i},
\nonumber\\
A_{\beta_i}^G
&=
A_{\beta_i}^P
+
\cp B_{\beta_i}\sinh(\lambda_i)\,G_{\alpha_i}
-
B_{\beta_i}\cosh(\lambda_i)\,G_{\beta_i}.
\end{align}
These equations show that the drift gauges modify the drift terms of the
phase-space variables $\alpha_i$ and $\beta_i$, while introducing stochastic
fluctuations in the weight $\Omega$.

The trajectories generated by these SDEs sample the phase-space
distribution $P(\vecbf{x},t)$ propagated according to
Eq.~\eqref{eq:gauged_fpe}.  For a phase-space function
$X(\vecbf{x})$, we define the weighted average as
\begin{align}\label{eq:weighted_ensemble_average}
\wexp{X}
\equiv
\frac{
\int d\mu\,P(\vecbf{x},t)\Omega X(\vecbf{x})
}{
\int d\mu\,P(\vecbf{x},t)\Omega
}
\simeq
\frac{1}{N_w}
\sum_{w=1}^{N_w}
\widetilde\Omega_w(t)X_w(t),
\end{align}
where $w=1,\ldots,N_w$ labels the stochastic trajectories, and
\begin{align}
\widetilde\Omega_w(t)
=
\frac{N_w\Omega_w(t)}
{\sum_{w'=1}^{N_w}\Omega_{w'}(t)}
\end{align}
is the normalized trajectory weight.  For example,
$\langle\hat n_i\rangle=\wexp{\beta_i\alpha_i}$ and
$\langle\hat a_i^{\dagger 2}\hat a_i^2\rangle
=\wexp{\beta_i^2\alpha_i^2}$. This provides a stochastic sampling framework for the target Lindblad
dynamics under the vanishing-boundary-term assumption.
The positive-$P$ representation is recovered by setting
$\vecbf G=\vecbf0$, $\vecbf\lambda=\vecbf0$, and $\Omega=1$.

We parameterize the drift and diffusion gauges with a graph neural network,
which allows them to adapt flexibly as the phase-space configuration
evolves:
\begin{align}\label{eq:neural_gauge_map}
\left(
\vecbf G_\theta,
\vecbf\lambda_\theta
\right)
=
\mathcal N_\theta
\left(
\vecbf\alpha,
\vecbf\beta,
\ln\Omega,
t;\vecbf J
\right).
\end{align}
The network parameters $\theta$ are shared across all lattice sites, while
message passing along the hopping graph incorporates the connectivity
encoded by $\vecbf J$.  At each site $i$, the network outputs the two
complex drift gauges $G_{\alpha_i}$ and $G_{\beta_i}$ and the real
diffusion gauge $\lambda_i$.  Details of the architecture are provided in
Appendix~\ref{app:neural_architecture}.
In the next subsection, we show how projected boundary terms enter the
exact moment equations and use the resulting residuals to define the
optimization objective for $\theta$.

%----------------------------------------------------------------------
\subsection{Projected boundary terms and exact moment equation residuals}\label{sec:model_and_method:neural_gauge_optimization}
%----------------------------------------------------------------------

Because direct evaluation of the operator-valued boundary term in
Eq.~\eqref{eq:operator_boundary_term} is generally challenging, we instead
probe it through projections onto selected normally ordered monomial
operators.  If $\widehat{\mathcal V}(t)$ vanishes, each projected boundary
term must also vanish.  We show below that the projected boundary term
associated with a given monomial operator appears as an additional
contribution to the exact equation of motion for the corresponding normally
ordered moment.  The resulting exact moment equation residuals therefore
quantify these projected boundary terms and provide a tractable objective
for optimizing the neural drift and diffusion gauges.

For a distribution propagated by the Fokker--Planck equation, the density
operator represented by this distribution evolves according to
\begin{align}\label{eq:fpe_density_evolution}
\partial_t\hat\rho(t)
&=
\int d\mu\,
\gaugePKernel\,
\mathcal L_{\rm FP}^G[P(\vecbf{x},t)]
\nonumber\\
&=
\mathcal L[\hat\rho(t)]
+
\widehat{\mathcal V}(t),
\end{align}
where the second equality follows from
Eq.~\eqref{eq:explicit_integration_by_parts}.   Thus,
$\widehat{\mathcal V}(t)$ is the difference between the density-operator
evolution generated by the Fokker--Planck equation and the target Lindblad
dynamics.

We consider the site-local normally ordered monomials
$\hat O_{m,n,i}=(\hat a_i^\dagger)^m\hat a_i^n$, where
$i=1,\ldots,M$ labels the lattice site and
$m,n\in\mathbb N_0=\{0,1,2,\ldots\}$.  The corresponding normalized
moments are
$\mathcal O_{m,n,i}(t)\equiv
\langle\hat O_{m,n,i}\rangle
=\wexp{\beta_i^m\alpha_i^n}$.
Differentiating these moments and using the trace-preserving property
$\mytrace{\mathcal L[\hat\rho]}=0$ together with
$\mytrace{\hat O\,\mathcal L[\hat\rho]}
=
\mytrace{\mathcal L^\dagger[\hat O]\,\hat\rho}$
gives
\begin{align}\label{eq:projected_boundary_moment_equation}
\partial_t\mathcal O_{m,n,i}(t)
=
\mathcal F_{m,n,i}(t)
+
\mathcal V_{m,n,i}(t),
\end{align}
where
\begin{align}
\mathcal F_{m,n,i}(t)
&\equiv
\left\langle
\mathcal L^\dagger[\hat O_{m,n,i}]
\right\rangle,
\label{eq:exact_moment_rhs_definition}
\\
\mathcal V_{m,n,i}(t)
&\equiv
\frac{
\mytrace{
\hat O_{m,n,i}\widehat{\mathcal V}(t)
}
-
\mathcal O_{m,n,i}(t)
\mytrace{
\widehat{\mathcal V}(t)
}
}{
\mytrace{\hat\rho(t)}
}.
\label{eq:normalized_projected_boundary}
\end{align}
The subtraction proportional to
$\mytrace{\widehat{\mathcal V}(t)}$ results from the normalization by
$\mytrace{\hat\rho(t)}$.  The Lindblad dynamics requires
$\partial_t\mathcal O_{m,n,i}=\mathcal F_{m,n,i}$; the additional term
$\mathcal V_{m,n,i}$ is the corresponding projection of the boundary term.

For the model in Eq.~\eqref{eq:lindblad}, the exact moment equation
right-hand side is
\begin{align}\label{eq:gauge_p_exact_moment_rhs}
\mathcal F_{m,n,i}
={}&
\Bigg[
\cp(n-m)\Delta
-\frac{\gamma}{2}(m+n)
\nonumber\\[-0.3ex]
&\qquad
+\frac{\cp U}{2}
\left(
m(m-1)-n(n-1)
\right)
\Bigg]
\wexp{\beta_i^m\alpha_i^n}
\nonumber\\
&+
\cp U(m-n)
\wexp{
\beta_i^{m+1}\alpha_i^{n+1}
}
\nonumber\\
&-
\cp nF
\wexp{
\beta_i^m\alpha_i^{n-1}
}
+
\cp mF^*
\wexp{
\beta_i^{m-1}\alpha_i^n
}
\nonumber\\
&+
\cp n
\sum_{\ell=1}^{M}
J_{i\ell}
\wexp{
\beta_i^m
\alpha_i^{n-1}
\alpha_\ell
}
\nonumber\\
&-
\cp m
\sum_{\ell=1}^{M}
J_{i\ell}^*
\wexp{
\beta_\ell
\beta_i^{m-1}
\alpha_i^n
}.
\end{align}
Terms containing negative powers are omitted.  All local and intersite
moments on the right-hand side are evaluated from the same trajectory
ensemble, and therefore no moment closure is introduced.

To avoid estimating the time derivative directly, we integrate the exact
moment equation over a time window $[t_k,t_{k+1}]$, with
$\Delta T_k=t_{k+1}-t_k$, and define the exact moment equation residual
\begin{align}\label{eq:window_exact_moment_residual}
\mathcal R_{k,m,n,i}
\equiv{}&
\frac{1}{\Delta T_k}
\Bigg[
\mathcal{O}_{m,n,i}(t_{k+1})
-
\mathcal{O}_{m,n,i}(t_k)
\nonumber\\
&\qquad
-
\int_{t_k}^{t_{k+1}}
dt\,
\mathcal F_{m,n,i}(t)
\Bigg]
\nonumber\\
={}&
\frac{1}{\Delta T_k}
\int_{t_k}^{t_{k+1}}
dt\,
\mathcal V_{m,n,i}(t).
\end{align}
Thus, $\mathcal R_{k,m,n,i}$ is the window average of the projected boundary
term in channel $(m,n,i)$.

For a finite ensemble of $N_w$ trajectories, we define the
trajectory-resolved contribution to $\mathcal O_{m,n,i}(t)$ as
\begin{align}
O_{w,m,n,i}(t)
=
\widetilde\Omega_w(t)\,
\beta_{w,i}^m(t)\alpha_{w,i}^n(t),
\end{align}
such that
\begin{align}
\mathcal O_{m,n,i}(t)
\simeq
\frac{1}{N_w}
\sum_{w=1}^{N_w}
O_{w,m,n,i}(t).
\end{align}
Similarly, we denote by $F_{w,m,n,i}(t)$ the trajectory-resolved contribution
to $\mathcal F_{m,n,i}(t)$, such that
\begin{align}
\mathcal F_{m,n,i}(t)
\simeq
\frac{1}{N_w}
\sum_{w=1}^{N_w}
F_{w,m,n,i}(t).
\end{align}
The corresponding trajectory-resolved contribution to the exact moment
equation residual over the time window $[t_k,t_{k+1}]$ is then
\begin{align}\label{eq:trajectory_exact_moment_residual}
R_{k,w,m,n,i}
\equiv{}&
\frac{1}{\Delta T_k}
\Bigg[
O_{w,m,n,i}(t_{k+1})
-
O_{w,m,n,i}(t_k)
\nonumber\\
&\qquad
-
\int_{t_k}^{t_{k+1}}
dt\,
F_{w,m,n,i}(t)
\Bigg].
\end{align}
For a selected set of monomial channels,
\begin{align}
\mathcal C
=
\left\{
(m_a,n_a)
\right\}_{a=1}^{N_{\mathcal C}},
\end{align}
we assemble the real and imaginary parts of the corresponding
trajectory-resolved residual contributions into the vector
\begin{align}
\vecbf R_{k,w,i}
=
\Bigl(
&\operatorname{Re}R_{k,w,m_1,n_1,i},
\operatorname{Im}R_{k,w,m_1,n_1,i},
\nonumber\\
&\ldots,
\operatorname{Re}R_{k,w,m_{N_{\mathcal C}},n_{N_{\mathcal C}},i},
\operatorname{Im}R_{k,w,m_{N_{\mathcal C}},n_{N_{\mathcal C}},i}
\Bigr)^T.
\end{align}
The sample mean and covariance of these residual vectors are
\begin{align}\label{eq:residual_mean_covariance}
\vecbf\mu_{k,i}
&=
\frac{1}{N_w}
\sum_{w=1}^{N_w}
\vecbf R_{k,w,i},
\nonumber\\
\vecbf\Sigma_{k,i}
&=
\frac{1}{N_w-1}
\sum_{w=1}^{N_w}
\left(
\vecbf R_{k,w,i}
-
\vecbf\mu_{k,i}
\right)
\left(
\vecbf R_{k,w,i}
-
\vecbf\mu_{k,i}
\right)^T.
\end{align}

The exact moment equation residuals $\mathcal R_{k,m,n,i}$ associated with different monomial
channels can differ substantially in their natural scales and sampling
fluctuations.  Because all channels are
evaluated from the same trajectory ensemble, their residuals are also
correlated.  Consequently, an unweighted sum of squared residuals can be
dominated by channels with large magnitudes or strong fluctuations.  We therefore use the joint residual covariance as a preconditioner and
define
\begin{align}\label{eq:cov_exact_moment_loss}
z_{k,i}^2
&=
\vecbf\mu_{k,i}^{T}
\operatorname{sg}
\left[
\left(
\vecbf\Sigma_{k,i}
+
\epsilon\vecbf I
\right)^{-1}
\right]
\vecbf\mu_{k,i},
\nonumber\\
\mathrm{Loss}
&=
\frac{1}{N_{\rm win}}
\sum_{k=1}^{N_{\rm win}}
\frac{1}{M}
\sum_{i=1}^{M}
z_{k,i}^2.
\end{align}
Here $\epsilon>0$ is a small diagonal regularizer, and
$\operatorname{sg}[\cdot]$ denotes stop-gradient, so $\vecbf\Sigma_{k,i}$
sets the relative scales and correlations of the residual channels without
being directly optimized.  $N_{\rm win}$ denotes the number
of time windows; for windows of equal duration $\Delta T$, the total
training time is $t_{\mathrm{train}}=N_{\rm win}\Delta T$.

The choice of $\mathcal C$ balances the set of projected boundary terms
included in the optimization objective against the sampling quality of the corresponding
exact moment equations.  Including higher-order monomial channels probes
higher-order structure in the sampled phase-space distribution and thereby
imposes stronger moment constraints on it.  The associated normally ordered
moments, however, are more difficult to estimate accurately and generally
exhibit larger sampling fluctuations, which can obscure the exact moment
equation residuals and impede optimization.

In this work, we choose
\begin{align}\label{eq:default_monomial_set}
\mathcal C
=
\left\{
(0,1),
(1,0),
(0,2),
(1,1),
(2,0),
(2,2)
\right\}.
\end{align}
These channels correspond to the first-order field moments, the anomalous
pair moments, the local density, and the fourth-order moment required to
evaluate the local second-order correlation.  The diagonal channels are
particularly useful because, for $m=n$, both
terms generated by the onsite Kerr interaction $U$ in Eq.~\eqref{eq:gauge_p_exact_moment_rhs} vanish.  Their exact moment equations therefore do not involve moments of order
$m+n+2$ and can exhibit smaller sampling fluctuations than off-diagonal
channels of the same total order.  The $(1,1)$ and $(2,2)$ channels thus
provide higher-order constraints with comparatively small sampling
fluctuations.

%----------------------------------------------------------------------
\section{Results and discussion}\label{sec:results}
%----------------------------------------------------------------------

In this section, we present results obtained with the neural gauge-$P$
representation.  We first consider the undriven single-site model and
analyze the learned gauge fields.  We then study the coherently driven
single-site model in the antibunching regime, where the convergence with trajectory number is examined. Finally, we apply the method to a $16\times16$ square lattice
and compare the long-time results with corner-space renormalization method steady-state values.  In all calculations, the neural gauges are trained using
$N_w=6000$ stochastic trajectories with time step
$dtU=10^{-3}$ and time window length $\Delta T U = 0.1$, whereas subsequent simulations with the trained
gauges use a smaller time step of $dtU=10^{-4}$. For both training and simulation, the neural gauges are applied every $10$ time steps. 

\subsection{Single-site model without driving}
%----------------------------------------------------------------------
% Fig 2
%----------------------------------------------------------------------
\begin{figure}[!t]
    \centering
    \includegraphics[width=\columnwidth]{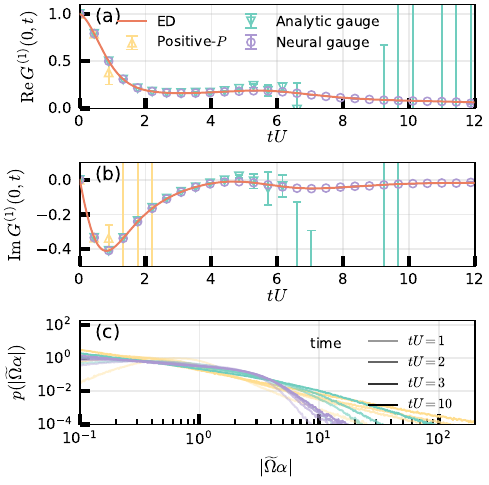}
    \caption{
        Single-site Kerr dynamics for $U=1$, $\gamma=0.3U$, $F=0$,
        $\Delta=0$, initialized with $\alpha(0)=\beta(0)=\Omega(0)=1$.
        The neural gauges are
        trained up to $t_{\mathrm{train}}U=8$ using a GNN with one
        message-passing layer and hidden dimension $h=256$.  For
        positive-$P$, gauge-$P$ with the analytic gauge, and neural
        gauge-$P$, the observables are estimated from $N_w=10^6$
        stochastic trajectories.  Panels (a) and (b) show the real and
        imaginary parts of
        $G^{(1)}(0,t)$ from exact
        diagonalization (red solid line), positive-$P$ (yellow upward
        triangles), gauge-$P$ with the analytic gauge (green downward
        triangles), and neural gauge-$P$ (purple circles); error bars
        denote standard errors.  Panel (c) shows the empirical
        distributions of $|\widetilde\Omega\alpha|$ at
        $tU=1,2,3$, and $10$ for the three phase-space representations,
        using the same colors as in panels (a) and (b) and increasing
        opacity at later times.
    }
\label{fig:single_advantage}
\end{figure}
We first present the single-site Kerr dynamics in
Fig.~\ref{fig:single_advantage} for $U=1$, $\gamma=0.3U$, $F=0$,
$\Delta=0$, initialized with
$\alpha(0)=\beta(0)=\Omega(0)=1$, corresponding to a coherent initial
state with mean occupation $n(0)=1$.  Panels (a) and (b) show,
respectively, the real and imaginary parts of the first-order temporal
correlation function
$G^{(1)}(0,t)=\langle\hat a^\dagger(0)\hat a(t)\rangle$.
At short times, $tU\lesssim0.7$, the results obtained with positive-$P$,
gauge-$P$ using the analytic gauge of
Ref.~\cite{Wuster_positive_P_and_gauge_P_2017}, and neural gauge-$P$ all
agree well with exact diagonalization.  At later times, the positive-$P$
result develops large fluctuations and spikes.
Gauge-$P$ with the analytic gauge postpones the onset of this spiking
behavior to $tU\simeq6$, whereas neural gauge-$P$ remains in agreement
with exact diagonalization over a substantially longer time interval,
extending toward the steady state, even though the neural gauges are
trained only up to $t_{\mathrm{train}}U=8$.

Figure~\ref{fig:single_advantage}(c) depicts the corresponding empirical
distributions of the weighted-estimator magnitude
$|\widetilde\Omega\alpha|$.  For positive-$P$ and gauge-$P$ with the
analytic gauge, the distributions broaden with time and develop slowly
decaying tails.  Such tails can produce nonvanishing boundary terms and,
in a finite trajectory ensemble, give rise to rare trajectories with
exceptionally large weighted contributions.  These trajectories can then
dominate the ensemble average and generate the spikes observed in
panels (a) and (b).  By contrast, neural gauge-$P$ maintains a compact
distribution and suppresses these rare large contributions, consistent
with the improved long-time accuracy of $G^{(1)}(0,t)$.

%----------------------------------------------------------------------
% Fig. 3
%----------------------------------------------------------------------
\begin{figure}[!t]
    \centering
    \includegraphics[width=\columnwidth]{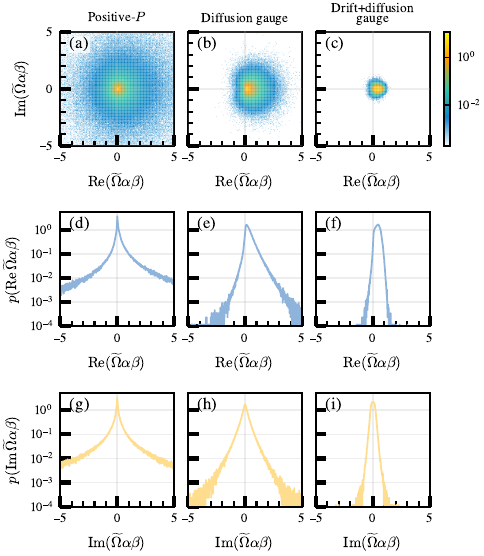}
    \caption{
        Empirical distributions of the weighted occupation estimator
        $\widetilde\Omega\alpha\beta$, estimated from stochastic trajectories at
        $tU=3$ for the undriven single-site system of
        Fig.~\ref{fig:single_advantage}.  Panels (a)--(c) show the joint empirical
        distributions in the complex plane for positive-$P$, neural gauge-$P$ with
        only the learned diffusion gauge retained, and full neural gauge-$P$ with
        both learned drift and diffusion gauges, respectively.  Panels (d)--(f)
        and (g)--(i) show the corresponding marginal distributions of the real and
        imaginary parts, respectively.
    }
    \label{fig:single_gauge_analysis}
\end{figure}
To identify the respective roles of the drift and diffusion gauges in
extending the useful simulation time demonstrated in
Fig.~\ref{fig:single_advantage}, we compare the empirical distributions of
the weighted occupation estimator $\widetilde\Omega\alpha\beta$ sampled
from stochastic realizations of the corresponding SDEs at $tU=3$.  The
left, middle, and right columns of
Fig.~\ref{fig:single_gauge_analysis} show results obtained with
positive-$P$, neural gauge-$P$ retaining only the learned diffusion gauge,
and full neural gauge-$P$, respectively. 
The diffusion-only calculation narrows the weighted-estimator distribution
and partially suppresses the broad tails present in positive-$P$, as seen
by comparing the joint empirical distributions in
Figs.~\ref{fig:single_gauge_analysis}(a) and
\ref{fig:single_gauge_analysis}(b).  The resulting distribution, however, remains considerably broader than
that obtained with full neural gauge-$P$ in
Fig.~\ref{fig:single_gauge_analysis}(c), as is seen more clearly from the
marginal distributions in the lower panels.  Compact marginals for both
the real and imaginary parts of the weighted estimator are obtained only
when the drift and diffusion gauges act together.  The extended useful
simulation time in Fig.~\ref{fig:single_advantage} therefore arises from
the combined action of these two gauge freedoms.

%----------------------------------------------------------------------
% Fig. 4
%----------------------------------------------------------------------
\begin{figure}[!t]
    \centering
    \includegraphics[width=\columnwidth]{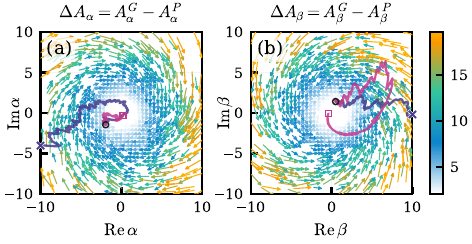}
    \caption{
        Gauge-induced drift fields for the single-site Kerr system of
        Fig.~\ref{fig:single_advantage} at $tU=3$.  Panels (a) and (b) show
        $\Delta A_{\alpha}=A_{\alpha}^{G}-A_{\alpha}^{P}$ and
        $\Delta A_{\beta}=A_{\beta}^{G}-A_{\beta}^{P}$, respectively,
        projected onto the complex $\alpha$ and $\beta$ planes.  The arrow length
        and color indicate the local field magnitude.  The blue and purple lines
        show trajectories generated by positive-$P$ and neural gauge-$P$,
        respectively.  Both trajectories start from the same initial condition and
        are propagated using the same Wiener-noise realization.  The open circle
        marks the initial point, the open squares mark the final points, and the
        cross marks where the positive-$P$ trajectory exits the displayed region.
    }
    \label{fig:gauge_direction}
\end{figure}
To gain further insight into how the learned gauges affect the stochastic
trajectories, we examine the gauge-induced drift fields in
Fig.~\ref{fig:gauge_direction}.
The fields $\Delta A_{\alpha}=A_{\alpha}^{G}-A_{\alpha}^{P}$ in
panel (a) and $\Delta A_{\beta}=A_{\beta}^{G}-A_{\beta}^{P}$ in
panel (b) are evaluated at phase-space points sampled from positive-$P$
trajectories.  As can be seen from Eq.~\eqref{eq:gauge_P_sde_components}, these fields
depend on both the learned drift gauges $\vecbf G_{\vecbf \theta}(\alpha, \beta, \Omega, t)$, and diffusion gauges $\vecbf \lambda_{\vecbf \theta}(\alpha, \beta, \Omega, t)$, and therefore reflect
their joint action on the stochastic dynamics. The
fields are negligible within the central region.  Away from this region,
their magnitude increases with the phase-space amplitude, while their
directions exhibit a pronounced rotational component and point inward
toward the center.  The combined gauges therefore generate a rotational
flow with a restoring radial component that becomes progressively stronger
for large phase-space excursions.  The effect of this restoring flow is
illustrated by the overlaid trajectories.  Starting from the same initial condition and
Wiener-noise realization, the positive-$P$ trajectory undergoes a large
excursion and leaves the displayed region, whereas the neural gauge-$P$
trajectory remains within the central region.  The learned gauges thus
reshape the trajectory flow smoothly rather than imposing a hard
constraint on the phase-space variables.  This restoring flow explains
the compact weighted-estimator distribution in
Fig.~\ref{fig:single_gauge_analysis} in terms of the underlying stochastic
phase-space dynamics.

\subsection{Single-site model with driving}
%----------------------------------------------------------------------
% Fig 5
%----------------------------------------------------------------------
\begin{figure}[!t]
    \centering
    \includegraphics[width=\columnwidth]{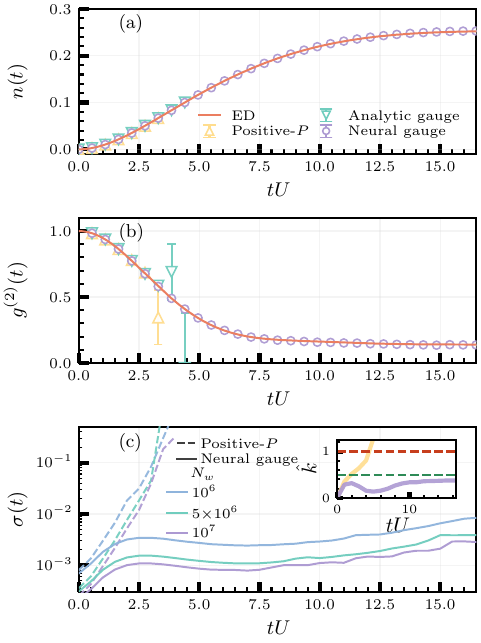}
    \caption{
            Coherently driven single-site dynamics for $U=1$, $\gamma=0.3U$,
            $\Delta=0$, and $F=0.1U$, initialized in the vacuum state.  The neural
            gauges are trained up to $Ut_{\mathrm{train}}=12$ using a GNN with one
            message-passing layer and hidden dimension $h=256$.  Panels (a) and (b)
            show $n(t)$ and $g^{(2)}(t)$, respectively, obtained from exact
            diagonalization (red solid line), positive-$P$ (yellow upward triangles),
            gauge-$P$ with the analytic gauge (green downward triangles), and neural
            gauge-$P$ (purple circles).  The stochastic estimates use $N_w=10^7$
            trajectories, and the error bars denote standard errors.  Panel (c) shows
            the standard error $\sigma(t)$ of the $g^{(2)}(t)$ estimator for
            $N_w=1\times10^6$, $5\times10^6$, and $10^7$.
            Solid and dashed lines denote neural gauge-$P$ and positive-$P$,
            respectively, with the same color indicating the same trajectory number.
            The inset shows the Pareto-$\hat{k}$ diagnostic for positive-$P$ and
            neural gauge-$P$, using the same colors and symbols as in panels (a) and
            (b).  The green and red horizontal dashed lines indicate $\hat{k}=0.5$
            and $\hat{k}=1$, respectively.
            }
\label{fig:single_samples_convergence}
\end{figure}
We next consider the coherently driven single-site case with $F=0.1U$,
retaining the remaining parameters of Fig.~\ref{fig:single_advantage} and
initializing the system in the vacuum state.  The resulting steady state
has a finite occupation and lies in the antibunching regime, with
$g^{(2)}(t\rightarrow\infty)=
\langle\hat a^{\dagger 2}(t)\hat a^2(t)\rangle/
\langle\hat n(t)\rangle^2<1$. Panels Fig.~\ref{fig:single_samples_convergence}(a) and Fig.~\ref{fig:single_samples_convergence}(b) show
$n(t)$ and $g^{(2)}(t)$, respectively, comparing results obtained with
positive-$P$, gauge-$P$ using the analytic gauge, and neural gauge-$P$
against exact diagonalization.  The neural gauge-$P$ results remain in
agreement with exact diagonalization throughout the long-time evolution
toward the steady state, whereas the positive-$P$ estimates develop spikes
and become unreliable around $tU \simeq 3.3$.  The analytic gauge slightly improves the stability of the positive-$P$ results but still fails around $tU\simeq 4$.  The agreement in $g^{(2)}(t)$ provides a more
stringent test because it depends on a fourth-order moment and is therefore
more sensitive to broad estimator tails. Nevertheless, neural gauge-$P$ accurately
captures both the antibunching dynamics and the subsequent approach to the
steady state.

Figure~\ref{fig:single_samples_convergence}(c) examines the sample-size
convergence of $g^{(2)}(t)$.  For neural gauge-$P$, increasing the number
of trajectories systematically reduces the estimated standard error $\sigma(t)$.  The
positive-$P$ results instead do not show significant improvement.  We attribute this contrasting
behavior to the heavy-tailed distribution of the trajectory contributions
in positive-$P$. To show this quantitatively, the inset presents
the corresponding Pareto-$\hat{k}$ diagnostics~\cite{Vehtari_Pareto_2024}.  For neural gauge-$P$,
$\hat{k}$ remains below $1/2$, consistent with a finite variance.  For positive-$P$, by contrast, $\hat{k}$ exceeds $1$ at early times,
indicating that the estimator does not possess a finite mean.

\subsection{Lattice model}
%----------------------------------------------------------------------
% Lattice system
%----------------------------------------------------------------------
\begin{figure}[!t]
    \centering
    \includegraphics[width=\columnwidth]{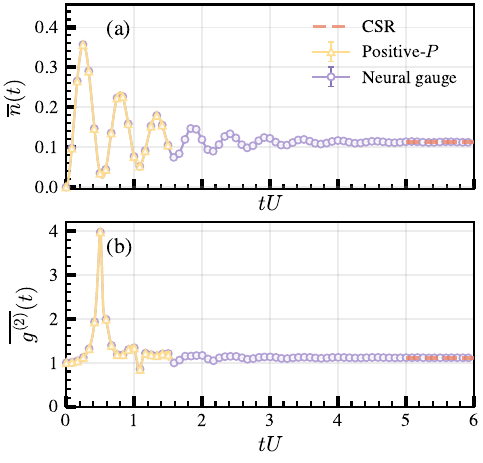}
    \caption{
    Driven-dissipative Bose--Hubbard model on a $16\times16$ square
    lattice with periodic boundary conditions, $U=0.5$, $J=0.5U$,
    $\gamma=2U$, $F=4U$, and $\Delta=10U$.  Panels (a) and (b) show the
    lattice-averaged density $\overline{n}(t)$ and normalized local
    second-order correlation $\overline{g^{(2)}}(t)$, respectively,
    obtained with positive-$P$ (open yellow triangles) and neural
    gauge-$P$ (open purple circles), together with
    corner-space-renormalization steady-state reference values (red dashed
    lines).  The neural gauges are trained up to
    $t_{\mathrm{train}}U=4$ using a GNN with one message-passing layer and
    hidden dimension $h=256$.  For both positive-$P$ and neural
    gauge-$P$, the observables are estimated from $N_w=10^6$ stochastic
    trajectories initialized in the vacuum state.
    }
    \label{fig:lattice_advantage}
\end{figure}
Having established the improved long-time performance of neural gauge-$P$
for the single-site cases, we next apply the method to the
driven-dissipative Bose--Hubbard model on a $16\times16$ square lattice.
Figure~\ref{fig:lattice_advantage} compares the positive-$P$ and neural
gauge-$P$ results with corner-space-renormalization steady-state reference
calculations~\cite{Finazzi_corner_space_2015} for $U=0.5$,
$J=0.5U$, $\gamma=2U$, $F=4U$, and $\Delta=10U$.  The neural gauges are
parameterized by the site-shared graph neural network introduced above.
At short times, the two phase-space representations give consistent
results for both the lattice-averaged density $\overline{n}(t)$ in
panel (a) and the lattice-averaged normalized local second-order
correlation $\overline{g^{(2)}}(t)$ in panel (b).  At later times, the
positive-$P$ estimates develop large fluctuations, whereas the neural
gauge-$P$ results remain well resolved and approach the
corner-space-renormalization steady-state values.  The agreement in
$\overline{g^{(2)}}(t)$ provides a more stringent test than that in
$\overline{n}(t)$ because it depends on a fourth-order moment and is
therefore more sensitive to broad estimator tails.

\section{Conclusion and outlook}
\label{sec:conclusion}

In this work, we introduce the neural gauge-$P$ representation for
interacting open bosonic systems.  The drift and diffusion gauges are
parameterized by neural networks and optimized using exact moment equation
residuals that quantify selected projected boundary terms, thereby linking
the optimization objective directly to the boundary terms that limit the
reliable simulation time of phase-space methods.  For the single-site
cases, both without and with coherent driving, the learned gauges maintain
compact weighted-estimator distributions and enable reliable long-time
evolution toward the steady state.  The gauge-induced drift fields show
that the drift and diffusion gauges act together to generate a rotational
flow with a restoring radial component that suppresses large phase-space
excursions.  For the square lattice, neural gauge-$P$ accurately captures
the density and local second-order correlation dynamics during long-time
evolution toward the steady state, whereas positive-$P$ ceases to yield
reliable results at substantially earlier times.

Further directions include extending the framework to nonlocal gauges that
couple stochastic degrees of freedom at different
sites~\cite{Wuster_positive_P_and_gauge_P_2017}, as well as to systems with
coherent two-particle driving and two-particle loss.  We expect the neural
gauge-$P$ representation to provide a valuable tool for studying
nonequilibrium open bosonic dynamics.

\begin{acknowledgments}
We thank Xinyang Dong and Yi Lu for stimulating discussions.  X. Cao
acknowledges support from the National Key R\&D Program of China under
Grant No.~2024YFA1408602. All results presented here were obtained using a single NVIDIA A800 GPU.
\end{acknowledgments}
%======================================================================
% APPENDICES
%======================================================================
\appendix
\section{Graph neural network architecture}
\label{app:neural_architecture}

In the calculations reported here, the drift and diffusion gauges are
parameterized by a graph neural network whose parameters are shared across
all lattice sites.  For each stochastic trajectory, the network receives
the instantaneous phase-space variables $\vecbf\alpha$ and
$\vecbf\beta$, the complex log weight $\ln\Omega$, and time $t$.  We use
$\ln\Omega$ rather than $\Omega$ as the network input for numerical
stability.  The hopping matrix $\vecbf J$ defines the graph along which
information is exchanged between sites.  At each site $i$, the network
outputs two complex drift gauges and one real diffusion gauge,
\begin{align}
G_{\alpha_i}
&=
G_{\alpha_i}^{\mathrm R}
+
\cp G_{\alpha_i}^{\mathrm I},
\nonumber\\
G_{\beta_i}
&=
G_{\beta_i}^{\mathrm R}
+
\cp G_{\beta_i}^{\mathrm I},
\nonumber\\
\lambda_i
&\in\mathbb R.
\end{align}

\paragraph{Input representation.}
Each nonzero matrix element $J_{ij}$ defines a directed edge from site
$j$ to site $i$, with $J_{ij}$ used as the corresponding edge weight.
The present implementation considers real hopping amplitudes.

The node features at site $i$ are constructed from
\begin{align}
\mathcal Z_i
=
\left\{
\alpha_i,\,
\beta_i,\,
n_i,\,
A_{\alpha_i}^{P},\,
A_{\beta_i}^{P}
\right\},
\qquad
n_i=\alpha_i\beta_i.
\end{align}
For each complex variable $z\in\mathcal Z_i$, we use the feature map
\begin{align}
\varphi(z)
=
\left(
\operatorname{Re}z,\,
\operatorname{Im}z
\right).
\end{align}
The complete node-feature vector is
\begin{align}
\vecbf q_i
=
\operatorname{concat}
\left[
\varphi(\alpha_i),
\varphi(\beta_i),
\varphi(n_i),
\varphi(A_{\alpha_i}^{P}),
\varphi(A_{\beta_i}^{P}),
d_i
\right],
\end{align}
where
\begin{align}
    d_i
    =
    \frac{
    \sum_j|J_{ij}|
    }{
    \max\left(1,\max_k\sum_j|J_{kj}|\right)
    },
\end{align}
is the normalized hopping degree.

The positive-$P$ drift features $A_{\alpha_i}^{P}$ and
$A_{\beta_i}^{P}$ are evaluated from Eq.~\eqref{eq:p_drift}.  They encode
the local deterministic dynamics, including the hopping fields
$(\vecbf J\vecbf\alpha)_i$ and
$(\vecbf J^{T}\vecbf\beta)_i$ generated by the sites coupled to $i$.
These physics-informed features allow each site to access the surrounding
lattice dynamics without requiring multiple message-passing layers to
reconstruct it from the phase-space coordinates alone.

The trajectory-dependent log weight and time are incorporated through the
global feature vector
\begin{align}
\vecbf u(t)
=
\operatorname{concat}
\left[
\operatorname{Re}\ln\Omega,\,
\operatorname{Im}\ln\Omega,\,
t
\right].
\end{align}

\paragraph{Encoder and global conditioning.}
The node and global features are embedded separately.  The global feature
vector is mapped to a hidden conditioning vector $\vecbf g$ and then
projected to the node space as $\vecbf c$, while the node features are
mapped to the initial hidden states:
\begin{align}
\vecbf g
&=
\sigma
\left(
\vecbf W_g\vecbf u+\vecbf b_g
\right),
\nonumber\\
\vecbf c
&=
\sigma
\left(
\vecbf W_c\vecbf g+\vecbf b_c
\right),
\nonumber\\
\vecbf h_i^{(0)}
&=
\sigma
\left(
\operatorname{LN}
\left(
\vecbf W_{\rm enc}\vecbf q_i+\vecbf b_{\rm enc}
\right)
+
\vecbf c
\right).
\end{align}
Here $\sigma$ denotes the GELU activation and
$\operatorname{LN}$ denotes layer normalization.  The conditioning vector
$\vecbf g$ is used to modulate the message-passing layers, while
$\vecbf c$ is added to the hidden representation at every site.

In all calculations reported in this work, the node embedding, message,
local-update, and global-conditioning representations use the same hidden
dimension $h$.  The graph processor contains $N_\ell$ message-passing
layers, so the GNN architecture used in this work is specified by the two
hyperparameters $(N_\ell,h)$.

\paragraph{Message-passing processor.}
The global conditioning modulates the hidden states through feature-wise
linear modulation,
\begin{align}
\operatorname{FiLM}_{\ell}
\left(
\vecbf \xi;\vecbf g
\right)
=
\vecbf \xi
\odot
\left[
1+
\tanh
\vecbf s_{\ell}(\vecbf g)
\right]
+
\vecbf b_{\ell}(\vecbf g),
\end{align}
where $\vecbf s_{\ell}$ and $\vecbf b_{\ell}$ are linear projections of
$\vecbf g$, and $\odot$ denotes elementwise multiplication.

Each layer $\ell=0,\ldots,N_\ell-1$ consists of a graph-message update
followed by a site-local residual update.  The conditioned hidden state is
first defined as
\begin{align}
\vecbf \xi_i^{(\ell)}
=
\operatorname{FiLM}_{\ell}
\left(
\operatorname{LN}
\left[
\vecbf h_i^{(\ell)}
+
\vecbf c
\right];
\vecbf g
\right).
\end{align}
The hopping-weighted message received by site $i$ is
\begin{align}
\vecbf m_i^{(\ell)}
=
\sum_j
J_{ij}
\vecbf W_{\rm in}^{(\ell)}
\vecbf \xi_j^{(\ell)}.
\end{align}
The message and local hidden state are then combined through the residual
update
\begin{align}
\widetilde{\vecbf h}_i^{(\ell)}
=
\vecbf h_i^{(\ell)}
+
\sigma
\left[
\vecbf W_{\rm self}^{(\ell)}
\vecbf x_i^{(\ell)}
+
\vecbf W_{\rm msg}^{(\ell)}
\vecbf m_i^{(\ell)}
\right].
\end{align}
A second conditioned residual update acts locally:
\begin{align}
\vecbf y_i^{(\ell)}
&=
\operatorname{FiLM}_{\ell}
\left(
\operatorname{LN}
\left[
\widetilde{\vecbf h}_i^{(\ell)}
+
\vecbf c
\right];
\vecbf g
\right),
\nonumber\\
\vecbf h_i^{(\ell+1)}
&=
\widetilde{\vecbf h}_i^{(\ell)}
+
\sigma
\left[
\vecbf W_{\rm loc}^{(\ell)}
\vecbf y_i^{(\ell)}
\right].
\end{align}
The first residual update propagates information through the hopping
graph, while the second refines the local hidden representation after
message aggregation.

All trainable parameters are shared across lattice sites, and the lattice
geometry enters through the hopping graph.  Consequently, for fixed
$(N_\ell,h)$, the number of network parameters is independent of the
number of sites.  For a single-site system, the graph message
$\vecbf m_i^{(\ell)}$ vanishes, and the architecture reduces to
$N_\ell$ globally conditioned local residual layers.

\paragraph{Output heads.}
After the final message-passing layer, the hidden state is transformed as
\begin{align}
\overline{\vecbf h}_i
=
\sigma
\left[
\operatorname{LN}
\left(
\vecbf h_i^{(N_\ell)}
\right)
\right].
\end{align}
Three site-wise linear heads produce the gauge outputs:
\begin{align}
\begin{pmatrix}
G_{\alpha_i}^{\mathrm R}\\
G_{\alpha_i}^{\mathrm I}
\end{pmatrix}
&=
\vecbf W_{\alpha}
\overline{\vecbf h}_i
+
\vecbf b_{\alpha},
\nonumber\\
\begin{pmatrix}
G_{\beta_i}^{\mathrm R}\\
G_{\beta_i}^{\mathrm I}
\end{pmatrix}
&=
\vecbf W_{\beta}
\overline{\vecbf h}_i
+
\vecbf b_{\beta},
\nonumber\\
\lambda_i
&=
\vecbf W_{\lambda}
\overline{\vecbf h}_i
+
b_{\lambda}.
\end{align}
The first two heads provide the real and imaginary parts of the complex
drift gauges $G_{\alpha_i}$ and $G_{\beta_i}$, while the third produces
the real diffusion gauge $\lambda_i$.

\bibliography{ngpsr}

@article{Lindblad_generators_1976,
  title={On the generators of quantum dynamical semigroups},
  author={Lindblad, Goran},
  journal={Communications in mathematical physics},
  volume={48},
  number={2},
  pages={119--130},
  year={1976},
  doi={https://doi.org/10.1007/BF01608499},
  publisher={Springer}
}

@article{Gorini_completely_positive_1976,
  title={Completely positive dynamical semigroups of N-level systems},
  author={Gorini, Vittorio and Kossakowski, Andrzej and Sudarshan, Ennackal Chandy George},
  journal={Journal of Mathematical Physics},
  volume={17},
  number={5},
  pages={821--825},
  year={1976},
  doi={https://doi.org/10.1063/1.522979},
  publisher={American Institute of Physics}
}

@article{Diehl_open_cold_atoms_2008,
  title={Quantum states and phases in driven open quantum systems with cold atoms},
  author={Diehl, Sebastian and Micheli, Andrea and Kantian, Adrian and Kraus, B and B{\"u}chler, Hans Peter and Zoller, Peter},
  journal={Nature Physics},
  volume={4},
  number={11},
  pages={878--883},
  year={2008},
  doi={https://doi.org/10.1038/nphys1073},
  publisher={Nature Publishing Group UK London}
}

@article{Sieberer_Keldysh_open_quantum_systems_2016,
doi = {10.1088/0034-4885/79/9/096001},
url = {https://doi.org/10.1088/0034-4885/79/9/096001},
year = {2016},
month = {aug},
publisher = {IOP Publishing},
volume = {79},
number = {9},
pages = {096001},
author = {Sieberer, L M and Buchhold, M and Diehl, S},
title = {Keldysh field theory for driven open quantum systems},
journal = {Reports on Progress in Physics}
}

@article{Minganti_spectral_theory_Liouvillians_2018,
  title = {Spectral theory of Liouvillians for dissipative phase transitions},
  author = {Minganti, Fabrizio and Biella, Alberto and Bartolo, Nicola and Ciuti, Cristiano},
  journal = {Phys. Rev. A},
  volume = {98},
  issue = {4},
  pages = {042118},
  numpages = {13},
  year = {2018},
  month = {Oct},
  publisher = {American Physical Society},
  doi = {10.1103/PhysRevA.98.042118},
  url = {https://link.aps.org/doi/10.1103/PhysRevA.98.042118}
}

@article{Macieszczak_metastability_open_quantum_dynamics_2016,
  title = {Towards a Theory of Metastability in Open Quantum Dynamics},
  author = {Macieszczak, Katarzyna and Gu\ifmmode \mbox{\c{t}}\else \c{t}\fi{}\ifmmode \u{a}\else \u{a}\fi{}, M\ifmmode \u{a}\else \u{a}\fi{}d\ifmmode \u{a}\else \u{a}\fi{}lin and Lesanovsky, Igor and Garrahan, Juan P.},
  journal = {Phys. Rev. Lett.},
  volume = {116},
  issue = {24},
  pages = {240404},
  numpages = {6},
  year = {2016},
  month = {Jun},
  publisher = {American Physical Society},
  doi = {10.1103/PhysRevLett.116.240404},
  url = {https://link.aps.org/doi/10.1103/PhysRevLett.116.240404}
}

@article{Sieberer_dynamical_critical_phenomena_2013,
  title = {Dynamical Critical Phenomena in Driven-Dissipative Systems},
  author = {Sieberer, L. M. and Huber, S. D. and Altman, E. and Diehl, S.},
  journal = {Phys. Rev. Lett.},
  volume = {110},
  issue = {19},
  pages = {195301},
  numpages = {5},
  year = {2013},
  month = {May},
  publisher = {American Physical Society},
  doi = {10.1103/PhysRevLett.110.195301},
  url = {https://link.aps.org/doi/10.1103/PhysRevLett.110.195301}
}

@article{verstraete_states_engineering_2009,
  title={Quantum computation and quantum-state engineering driven by dissipation},
  author={Verstraete, Frank and Wolf, Michael M and Ignacio Cirac, J},
  journal={Nature physics},
  volume={5},
  number={9},
  pages={633--636},
  year={2009},
  doi={https://doi.org/10.1038/nphys1342},
  publisher={Nature Publishing Group UK London}
}

@article{Carusotto_quantum_fluids_of_light_2013,
  title = {Quantum fluids of light},
  author = {Carusotto, Iacopo and Ciuti, Cristiano},
  journal = {Rev. Mod. Phys.},
  volume = {85},
  issue = {1},
  pages = {299--366},
  numpages = {0},
  year = {2013},
  month = {Feb},
  publisher = {American Physical Society},
  doi = {10.1103/RevModPhys.85.299},
  url = {https://link.aps.org/doi/10.1103/RevModPhys.85.299}
}

@article{Raimond_manipulating_2001,
  title = {Manipulating quantum entanglement with atoms and photons in a cavity},
  author = {Raimond, J. M. and Brune, M. and Haroche, S.},
  journal = {Rev. Mod. Phys.},
  volume = {73},
  issue = {3},
  pages = {565--582},
  numpages = {0},
  year = {2001},
  month = {Aug},
  publisher = {American Physical Society},
  doi = {10.1103/RevModPhys.73.565},
  url = {https://link.aps.org/doi/10.1103/RevModPhys.73.565}
}

@article{Walther_cavity_quantum_electrodynamics_2006,
doi = {10.1088/0034-4885/69/5/R02},
url = {https://doi.org/10.1088/0034-4885/69/5/R02},
year = {2006},
month = {apr},
publisher = {},
volume = {69},
number = {5},
pages = {1325},
author = {Walther, Herbert and Varcoe, Benjamin T H and Englert, Berthold-Georg and Becker, Thomas},
title = {Cavity quantum electrodynamics},
journal = {Reports on Progress in Physics}
}

@article{Reiserer_cavity_network_2015,
  title = {Cavity-based quantum networks with single atoms and optical photons},
  author = {Reiserer, Andreas and Rempe, Gerhard},
  journal = {Rev. Mod. Phys.},
  volume = {87},
  issue = {4},
  pages = {1379--1418},
  numpages = {40},
  year = {2015},
  month = {Dec},
  publisher = {American Physical Society},
  doi = {10.1103/RevModPhys.87.1379},
  url = {https://link.aps.org/doi/10.1103/RevModPhys.87.1379}
}

@article{Schmidt_circuit_QED_lattices_2013,
author = {Schmidt, Sebastian and Koch, Jens},
title = {Circuit QED lattices: Towards quantum simulation with superconducting circuits},
journal = {Annalen der Physik},
volume = {525},
number = {6},
pages = {395-412},
doi = {https://doi.org/10.1002/andp.201200261},
url = {https://onlinelibrary.wiley.com/doi/abs/10.1002/andp.201200261},
year = {2013}
}

@article{Carusotto_circuit_QED_2020,
  title={Photonic materials in circuit quantum electrodynamics},
  author={Carusotto, Iacopo and Houck, Andrew A and Koll{\'a}r, Alicia J and Roushan, Pedram and Schuster, David I and Simon, Jonathan},
  journal={Nature Physics},
  volume={16},
  number={3},
  pages={268--279},
  year={2020},
  doi={https://doi.org/10.1038/s41567-020-0815-y},
  publisher={Nature Publishing Group UK London}
}

@article{Hopfield_excitons_1958,
  title = {Theory of the Contribution of Excitons to the Complex Dielectric Constant of Crystals},
  author = {Hopfield, J. J.},
  journal = {Phys. Rev.},
  volume = {112},
  issue = {5},
  pages = {1555--1567},
  numpages = {0},
  year = {1958},
  month = {Dec},
  publisher = {American Physical Society},
  doi = {10.1103/PhysRev.112.1555},
  url = {https://link.aps.org/doi/10.1103/PhysRev.112.1555}
}

@article{Kim_polariton_condensation_2011,
  title={Dynamical d-wave condensation of exciton--polaritons in a two-dimensional square-lattice potential},
  author={Kim, Na Young and Kusudo, Kenichiro and Wu, Congjun and Masumoto, Naoyuki and L{\"o}ffler, Andreas and H{\"o}fling, Sven and Kumada, Norio and Worschech, Lukas and Forchel, Alfred and Yamamoto, Yoshihisa},
  journal={Nature Physics},
  volume={7},
  number={9},
  pages={681--686},
  year={2011},
  doi={https://doi.org/10.1038/nphys2012},
  publisher={Nature Publishing Group UK London}
}

@article{Klembt_polariton_TI_2018,
  title={Exciton-polariton topological insulator},
  author={Klembt, Sebastian and Harder, TH and Egorov, OA and Winkler, K and Ge, R and Bandres, MA and Emmerling, M and Worschech, L and Liew, TCH and Segev, M and others},
  journal={Nature},
  volume={562},
  number={7728},
  pages={552--556},
  year={2018},
  doi={https://doi.org/10.1038/s41586-018-0601-5},
  publisher={Nature Publishing Group UK London}
}

@article{Whittaker_polariton_Lieb_lattice_2018,
  title = {Exciton Polaritons in a Two-Dimensional Lieb Lattice with Spin-Orbit Coupling},
  author = {Whittaker, C. E. and Cancellieri, E. and Walker, P. M. and Gulevich, D. R. and Schomerus, H. and Vaitiekus, D. and Royall, B. and Whittaker, D. M. and Clarke, E. and Iorsh, I. V. and Shelykh, I. A. and Skolnick, M. S. and Krizhanovskii, D. N.},
  journal = {Phys. Rev. Lett.},
  volume = {120},
  issue = {9},
  pages = {097401},
  numpages = {6},
  year = {2018},
  month = {Mar},
  publisher = {American Physical Society},
  doi = {10.1103/PhysRevLett.120.097401},
  url = {https://link.aps.org/doi/10.1103/PhysRevLett.120.097401}
}

@article{Goblot_polariton_flatband_2019,
  title = {Nonlinear Polariton Fluids in a Flatband Reveal Discrete Gap Solitons},
  author = {Goblot, V. and Rauer, B. and Vicentini, F. and Le Boit\'e, A. and Galopin, E. and Lema\^{\i}tre, A. and Le Gratiet, L. and Harouri, A. and Sagnes, I. and Ravets, S. and Ciuti, C. and Amo, A. and Bloch, J.},
  journal = {Phys. Rev. Lett.},
  volume = {123},
  issue = {11},
  pages = {113901},
  numpages = {6},
  year = {2019},
  month = {Sep},
  publisher = {American Physical Society},
  doi = {10.1103/PhysRevLett.123.113901},
  url = {https://link.aps.org/doi/10.1103/PhysRevLett.123.113901}
}

@article{Su_polariton_condensation_2020,
  title={Observation of exciton polariton condensation in a perovskite lattice at room temperature},
  author={Su, Rui and Ghosh, Sanjib and Wang, Jun and Liu, Sheng and Diederichs, Carole and Liew, Timothy CH and Xiong, Qihua},
  journal={Nature Physics},
  volume={16},
  number={3},
  pages={301--306},
  year={2020},
  doi={https://doi.org/10.1038/s41567-019-0764-5},
  publisher={Nature Publishing Group UK London}
}

@article{Mirrahimi_protected_cat_qubits_2014,
  title={Dynamically protected cat-qubits: a new paradigm for universal quantum computation},
  author={Mirrahimi, Mazyar and Leghtas, Zaki and Albert, Victor V and Touzard, Steven and Schoelkopf, Robert J and Jiang, Liang and Devoret, Michel H},
  journal={New Journal of Physics},
  volume={16},
  number={4},
  pages={045014},
  year={2014},
  doi={10.1088/1367-2630/16/4/045014},
  publisher={IOP Publishing}
}

@article{Leghtas_two_photon_loss_2015,
author = {Z. Leghtas  and S. Touzard  and I. M. Pop  and A. Kou  and B. Vlastakis  and A. Petrenko  and K. M. Sliwa  and A. Narla  and S. Shankar  and M. J. Hatridge  and M. Reagor  and L. Frunzio  and R. J. Schoelkopf  and M. Mirrahimi  and M. H. Devoret },
title = {Confining the state of light to a quantum manifold by engineered two-photon loss},
journal = {Science},
volume = {347},
number = {6224},
pages = {853-857},
year = {2015},
doi = {10.1126/science.aaa2085},
URL = {https://www.science.org/doi/abs/10.1126/science.aaa2085}
}

@article{Grimm_Kerr_cat_qubit_2020,
  title={Stabilization and operation of a Kerr-cat qubit},
  author={Grimm, Alexander and Frattini, Nicholas E and Puri, Shruti and Mundhada, Shantanu O and Touzard, Steven and Mirrahimi, Mazyar and Girvin, Steven M and Shankar, Shyam and Devoret, Michel H},
  journal={Nature},
  volume={584},
  number={7820},
  pages={205--209},
  year={2020},
  doi={https://doi.org/10.1038/s41586-020-2587-z},
  publisher={Nature Publishing Group UK London}
}

@article{Gautier_cat_qubit_2022,
  title = {Combined Dissipative and Hamiltonian Confinement of Cat Qubits},
  author = {Gautier, Ronan and Sarlette, Alain and Mirrahimi, Mazyar},
  journal = {PRX Quantum},
  volume = {3},
  issue = {2},
  pages = {020339},
  numpages = {26},
  year = {2022},
  month = {May},
  publisher = {American Physical Society},
  doi = {10.1103/PRXQuantum.3.020339},
  url = {https://link.aps.org/doi/10.1103/PRXQuantum.3.020339}
}

@article{Chamberland_concatenated_cat_codes_2022,
  title = {Building a Fault-Tolerant Quantum Computer Using Concatenated Cat Codes},
  author = {Chamberland, Christopher and Noh, Kyungjoo and Arrangoiz-Arriola, Patricio and Campbell, Earl T. and Hann, Connor T. and Iverson, Joseph and Putterman, Harald and Bohdanowicz, Thomas C. and Flammia, Steven T. and Keller, Andrew and Refael, Gil and Preskill, John and Jiang, Liang and Safavi-Naeini, Amir H. and Painter, Oskar and Brand\~ao, Fernando G.S.L.},
  journal = {PRX Quantum},
  volume = {3},
  issue = {1},
  pages = {010329},
  numpages = {117},
  year = {2022},
  month = {Feb},
  publisher = {American Physical Society},
  doi = {10.1103/PRXQuantum.3.010329},
  url = {https://link.aps.org/doi/10.1103/PRXQuantum.3.010329}
}

@article{Pan_phase_space_compression_2023,
  title = {Protecting the Quantum Interference of Cat States by Phase-Space Compression},
  author = {Pan, Xiaozhou and Schwinger, Jonathan and Huang, Ni-Ni and Song, Pengtao and Chua, Weipin and Hanamura, Fumiya and Joshi, Atharv and Valadares, Fernando and Filip, Radim and Gao, Yvonne Y.},
  journal = {Phys. Rev. X},
  volume = {13},
  issue = {2},
  pages = {021004},
  numpages = {13},
  year = {2023},
  month = {Apr},
  publisher = {American Physical Society},
  doi = {10.1103/PhysRevX.13.021004},
  url = {https://link.aps.org/doi/10.1103/PhysRevX.13.021004}
}

@article{Hajr_Kerr_cat_qubit_2D_architecture_2024,
  title = {High-Coherence Kerr-Cat Qubit in 2D Architecture},
  author = {Hajr, Ahmed and Qing, Bingcheng and Wang, Ke and Koolstra, Gerwin and Pedramrazi, Zahra and Kang, Ziqi and Chen, Larry and Nguyen, Long B. and J\"unger, Christian and Goss, Noah and Huang, Irwin and Bhandari, Bibek and Frattini, Nicholas E. and Puri, Shruti and Dressel, Justin and Jordan, Andrew N. and Santiago, David I. and Siddiqi, Irfan},
  journal = {Phys. Rev. X},
  volume = {14},
  issue = {4},
  pages = {041049},
  numpages = {17},
  year = {2024},
  month = {Nov},
  publisher = {American Physical Society},
  doi = {10.1103/PhysRevX.14.041049},
  url = {https://link.aps.org/doi/10.1103/PhysRevX.14.041049}
}

@article{Hutin_neural_network_cat_states_2025,
  title = {Preparing Schr\"odinger Cat States in a Microwave Cavity Using a Neural Network},
  author = {Hutin, Hector and Bilous, Pavlo and Ye, Chengzhi and Abdollahi, Sepideh and Cros, Loris and Dvir, Tom and Shah, Tirth and Cohen, Yonatan and Bienfait, Audrey and Marquardt, Florian and Huard, Benjamin},
  journal = {PRX Quantum},
  volume = {6},
  issue = {1},
  pages = {010321},
  numpages = {22},
  year = {2025},
  month = {Jan},
  publisher = {American Physical Society},
  doi = {10.1103/PRXQuantum.6.010321},
  url = {https://link.aps.org/doi/10.1103/PRXQuantum.6.010321}
}

@article{Yuan_weak_Kerr_nonlinearities_2025,
  title = {Universal control in bosonic systems with weak Kerr nonlinearities},
  author = {Yuan, Ming and Seif, Alireza and Lingenfelter, Andrew and Schuster, David I. and Clerk, Aashish A. and Jiang, Liang},
  journal = {Phys. Rev. A},
  volume = {111},
  issue = {3},
  pages = {032606},
  numpages = {19},
  year = {2025},
  month = {Mar},
  publisher = {American Physical Society},
  doi = {10.1103/PhysRevA.111.032606},
  url = {https://link.aps.org/doi/10.1103/PhysRevA.111.032606}
}

@article{Ding_Kerr_cat_qubit_2025,
  title={Quantum control of an oscillator with a Kerr-cat qubit},
  author={Ding, Andy Z and Brock, Benjamin L and Eickbusch, Alec and Koottandavida, Akshay and Frattini, Nicholas E and Corti{\~n}as, Rodrigo G and Joshi, Vidul R and de Graaf, Stijn J and Chapman, Benjamin J and Ganjam, Suhas and others},
  journal={Nature Communications},
  volume={16},
  number={1},
  pages={5279},
  year={2025},
  doi={https://doi.org/10.1038/s41467-025-60352-w},
  publisher={Nature Publishing Group UK London}
}

@article{Verstraete_MPS_finite_temperature_dissipative_2004,
  title = {Matrix Product Density Operators: Simulation of Finite-Temperature and Dissipative Systems},
  author = {Verstraete, F. and Garc\'{\i}a-Ripoll, J. J. and Cirac, J. I.},
  journal = {Phys. Rev. Lett.},
  volume = {93},
  issue = {20},
  pages = {207204},
  numpages = {4},
  year = {2004},
  month = {Nov},
  publisher = {American Physical Society},
  doi = {10.1103/PhysRevLett.93.207204},
  url = {https://link.aps.org/doi/10.1103/PhysRevLett.93.207204}
}

@article{Cui_MPS_steady_state_2015,
  title = {Variational Matrix Product Operators for the Steady State of Dissipative Quantum Systems},
  author = {Cui, Jian and Cirac, J. Ignacio and Ba\~nuls, Mari Carmen},
  journal = {Phys. Rev. Lett.},
  volume = {114},
  issue = {22},
  pages = {220601},
  numpages = {5},
  year = {2015},
  month = {Jun},
  publisher = {American Physical Society},
  doi = {10.1103/PhysRevLett.114.220601},
  url = {https://link.aps.org/doi/10.1103/PhysRevLett.114.220601}
}

@article{Werner_positive_tensor_network_approach_2016,
  title = {Positive Tensor Network Approach for Simulating Open Quantum Many-Body Systems},
  author = {Werner, A. H. and Jaschke, D. and Silvi, P. and Kliesch, M. and Calarco, T. and Eisert, J. and Montangero, S.},
  journal = {Phys. Rev. Lett.},
  volume = {116},
  issue = {23},
  pages = {237201},
  numpages = {6},
  year = {2016},
  month = {Jun},
  publisher = {American Physical Society},
  doi = {10.1103/PhysRevLett.116.237201},
  url = {https://link.aps.org/doi/10.1103/PhysRevLett.116.237201}
}

@article{Kilda_projected_entangled_pair_operator_2021,
	title = {On the stability of the infinite Projected Entangled Pair Operator ansatz for driven-dissipative 2D lattices},
	pages = {005},
	author = {Kilda, Dainius and Biella, Alberto and Schirò, Marco and Fazio, Rosario and Keeling, Jonathan},
	journal = {SciPost Phys. Core},
	volume = {4},
	year = {2021},
	publisher = {SciPost},
	doi = {10.21468/SciPostPhysCore.4.1.005},
	url = {https://scipost.org/10.21468/SciPostPhysCore.4.1.005}
}

@article{Hryniuk_variational_Monte_Carlo_approach_2024,
  doi = {10.22331/q-2024-09-17-1475},
  url = {https://doi.org/10.22331/q-2024-09-17-1475},
  title = {Tensor-network-based variational {M}onte {C}arlo approach to the non-equilibrium steady state of open quantum systems},
  author = {Hryniuk, Dawid A. and Szyma{\'{n}}ska, Marzena H.},
  journal = {{Quantum}},
  issn = {2521-327X},
  publisher = {{Verein zur F{\"{o}}rderung des Open Access Publizierens in den Quantenwissenschaften}},
  volume = {8},
  pages = {1475},
  month = sep,
  year = {2024}
}

@article{Godinez_Riemannian_approach_2025,
  title = {Riemannian approach to the Lindbladian dynamics of a locally purified tensor network},
  author = {Godinez-Ramirez, Emiliano and Milbradt, Richard M. and Mendl, Christian B.},
  journal = {Phys. Rev. A},
  volume = {112},
  issue = {1},
  pages = {012224},
  numpages = {19},
  year = {2025},
  month = {Jul},
  publisher = {American Physical Society},
  doi = {10.1103/kztw-ywxr},
  url = {https://link.aps.org/doi/10.1103/kztw-ywxr}
}

@article{Sander_large_scale_stochastic_simulation_2025,
  title={Large-scale stochastic simulation of open quantum systems},
  author={Sander, Aaron and Fr{\"o}hlich, Maximilian and Eigel, Martin and Eisert, Jens and Gel{\ss}, Patrick and Hinterm{\"u}ller, Michael and Milbradt, Richard M and Wille, Robert and Mendl, Christian B},
  journal={Nature Communications},
  doi={https://doi.org/10.1038/s41467-025-66846-x},
  year={2025},
  publisher={Nature Publishing Group UK London}
}

@article{Torlai_neural_density_operators_2018,
  title = {Latent Space Purification via Neural Density Operators},
  author = {Torlai, Giacomo and Melko, Roger G.},
  journal = {Phys. Rev. Lett.},
  volume = {120},
  issue = {24},
  pages = {240503},
  numpages = {5},
  year = {2018},
  month = {Jun},
  publisher = {American Physical Society},
  doi = {10.1103/PhysRevLett.120.240503},
  url = {https://link.aps.org/doi/10.1103/PhysRevLett.120.240503}
}

@article{Hartmann_neural_network_approach_2019,
  title = {Neural-Network Approach to Dissipative Quantum Many-Body Dynamics},
  author = {Hartmann, Michael J. and Carleo, Giuseppe},
  journal = {Phys. Rev. Lett.},
  volume = {122},
  issue = {25},
  pages = {250502},
  numpages = {6},
  year = {2019},
  month = {Jun},
  publisher = {American Physical Society},
  doi = {10.1103/PhysRevLett.122.250502},
  url = {https://link.aps.org/doi/10.1103/PhysRevLett.122.250502}
}

@article{Vicentini_variational_neural_network_ansatz_2019,
  title = {Variational Neural-Network Ansatz for Steady States in Open Quantum Systems},
  author = {Vicentini, Filippo and Biella, Alberto and Regnault, Nicolas and Ciuti, Cristiano},
  journal = {Phys. Rev. Lett.},
  volume = {122},
  issue = {25},
  pages = {250503},
  numpages = {6},
  year = {2019},
  month = {Jun},
  publisher = {American Physical Society},
  doi = {10.1103/PhysRevLett.122.250503},
  url = {https://link.aps.org/doi/10.1103/PhysRevLett.122.250503}
}

@article{Nagy_vmc_neural_network_ansatz_2019,
  title = {Variational Quantum Monte Carlo Method with a Neural-Network Ansatz for Open Quantum Systems},
  author = {Nagy, Alexandra and Savona, Vincenzo},
  journal = {Phys. Rev. Lett.},
  volume = {122},
  issue = {25},
  pages = {250501},
  numpages = {6},
  year = {2019},
  month = {Jun},
  publisher = {American Physical Society},
  doi = {10.1103/PhysRevLett.122.250501},
  url = {https://link.aps.org/doi/10.1103/PhysRevLett.122.250501}
}

@article{Kothe_liouville_space_neural_network_representation_2024,
  title = {Liouville-space neural network representation of density matrices},
  author = {Kothe, Simon and Kirton, Peter},
  journal = {Phys. Rev. A},
  volume = {109},
  issue = {6},
  pages = {062215},
  numpages = {12},
  year = {2024},
  month = {Jun},
  publisher = {American Physical Society},
  doi = {10.1103/PhysRevA.109.062215},
  url = {https://link.aps.org/doi/10.1103/PhysRevA.109.062215}
}

@article{Mellak_neural_networks_variational_solutions_2024,
  title={Deep neural networks as variational solutions for correlated open quantum systems},
  author={Mellak, Johannes and Arrigoni, Enrico and von der Linden, Wolfgang},
  journal={Communications Physics},
  volume={7},
  number={1},
  pages={268},
  year={2024},
  doi={https://doi.org/10.1038/s42005-024-01757-9},
  publisher={Nature Publishing Group UK London}
}

@article{Lin_neural_density_operators_2024,
  title={Real-time dynamics of the Schwinger model as an open quantum system with Neural Density Operators},
  author={Lin, Joshua and Luo, Di and Yao, Xiaojun and Shanahan, Phiala E},
  journal={Journal of High Energy Physics},
  volume={2024},
  number={6},
  pages={1--23},
  year={2024},
  doi={https://doi.org/10.1007/JHEP06(2024)211},
  publisher={Springer}
}

@article{Denis_accurate_neural_quantum_states_2025,
  doi = {10.22331/q-2025-06-17-1772},
  url = {https://doi.org/10.22331/q-2025-06-17-1772},
  title = {Accurate neural quantum states for interacting lattice bosons},
  author = {Denis, Zakari and Carleo, Giuseppe},
  journal = {{Quantum}},
  issn = {2521-327X},
  publisher = {{Verein zur F{\"{o}}rderung des Open Access Publizierens in den Quantenwissenschaften}},
  volume = {9},
  pages = {1772},
  month = jun,
  year = {2025}
}

@article{Finazzi_corner_space_2015,
  title = {Corner-Space Renormalization Method for Driven-Dissipative Two-Dimensional Correlated Systems},
  author = {Finazzi, S. and Le Boit{\'e}, A. and Storme, F. and Baksic, A. and Ciuti, C.},
  journal = {Phys. Rev. Lett.},
  volume = {115},
  issue = {8},
  pages = {080604},
  year = {2015},
  month = {Aug},
  publisher = {American Physical Society},
  doi = {10.1103/PhysRevLett.115.080604},
  url = {https://link.aps.org/doi/10.1103/PhysRevLett.115.080604}
}

@article{Rota_quadratic_driven_2D_lattice_2019,
  title = {Quantum Critical Regime in a Quadratically Driven Nonlinear Photonic Lattice},
  author = {Rota, Riccardo and Minganti, Fabrizio and Ciuti, Cristiano and Savona, Vincenzo},
  journal = {Phys. Rev. Lett.},
  volume = {122},
  issue = {11},
  pages = {110405},
  numpages = {6},
  year = {2019},
  month = {Mar},
  publisher = {American Physical Society},
  doi = {10.1103/PhysRevLett.122.110405},
  url = {https://link.aps.org/doi/10.1103/PhysRevLett.122.110405}
}

@article{Drummond_positive_P_1980,
doi = {10.1088/0305-4470/13/7/018},
url = {https://doi.org/10.1088/0305-4470/13/7/018},
year = {1980},
month = {jul},
publisher = {},
volume = {13},
number = {7},
pages = {2353},
author = {P D Drummond and C W Gardiner},
title = {Generalised P-representations in quantum optics},
journal = {Journal of Physics A: Mathematical and General}
}

@article{Deuar_positive_P_2006,
doi = {10.1088/0305-4470/39/5/010},
url = {https://doi.org/10.1088/0305-4470/39/5/010},
year = {2006},
month = {jan},
publisher = {},
volume = {39},
number = {5},
pages = {1163},
author = {Deuar, P and Drummond, P D},
title = {First-principles quantum dynamics in interacting Bose gases: I. The positive P representation},
journal = {Journal of Physics A: Mathematical and General}
}

@article{Deuar_positive_P_2021,
  title = {Fully Quantum Scalable Description of Driven-Dissipative Lattice Models},
  author = {Deuar, Piotr and Ferrier, Alex and Matuszewski, Micha\l{} and Orso, Giuliano and Szyma\ifmmode \acute{n}\else \'{n}\fi{}ska, Marzena H.},
  journal = {PRX Quantum},
  volume = {2},
  issue = {1},
  pages = {010319},
  numpages = {19},
  year = {2021},
  month = {Feb},
  publisher = {American Physical Society},
  doi = {10.1103/PRXQuantum.2.010319},
  url = {https://link.aps.org/doi/10.1103/PRXQuantum.2.010319}
}

@article{Plimak_diffusion_gauge_2001,
  title = {Optimization of the positive-$P$ representation for the anharmonic oscillator},
  author = {Plimak, L. I. and Olsen, M. K. and Collett, M. J.},
  journal = {Phys. Rev. A},
  volume = {64},
  issue = {2},
  pages = {025801},
  numpages = {4},
  year = {2001},
  month = {Jul},
  publisher = {American Physical Society},
  doi = {10.1103/PhysRevA.64.025801},
  url = {https://link.aps.org/doi/10.1103/PhysRevA.64.025801}
}

@article{Deuar_gauge_P_2002,
  title = {Gauge $P$ representations for quantum-dynamical problems: Removal of boundary terms},
  author = {Deuar, P. and Drummond, P. D.},
  journal = {Phys. Rev. A},
  volume = {66},
  issue = {3},
  pages = {033812},
  numpages = {16},
  year = {2002},
  month = {Sep},
  publisher = {American Physical Society},
  doi = {10.1103/PhysRevA.66.033812},
  url = {https://link.aps.org/doi/10.1103/PhysRevA.66.033812}
}

@article{Deuar_gauge_P_2006,
doi = {10.1088/0305-4470/39/11/011},
url = {https://doi.org/10.1088/0305-4470/39/11/011},
year = {2006},
month = {mar},
publisher = {},
volume = {39},
number = {11},
pages = {2723},
author = {Deuar, P and Drummond, P D},
title = {First-principles quantum dynamics in interacting Bose gases II: stochastic gauges},
journal = {Journal of Physics A: Mathematical and General}
}

@article{Wuster_positive_P_and_gauge_P_2017,
  title = {Quantum dynamics of long-range interacting systems using the positive-$P$ and gauge-$P$ representations},
  author = {W\"uster, S. and Corney, J. F. and Rost, J. M. and Deuar, P.},
  journal = {Phys. Rev. E},
  volume = {96},
  issue = {1},
  pages = {013309},
  numpages = {22},
  year = {2017},
  month = {Jul},
  publisher = {American Physical Society},
  doi = {10.1103/PhysRevE.96.013309},
  url = {https://link.aps.org/doi/10.1103/PhysRevE.96.013309}
}

@article{Vehtari_Pareto_2024,
  author  = {Aki Vehtari and Daniel Simpson and Andrew Gelman and Yuling Yao and Jonah Gabry},
  title   = {Pareto Smoothed Importance Sampling},
  journal = {Journal of Machine Learning Research},
  year    = {2024},
  volume  = {25},
  number  = {72},
  pages   = {1--58},
  url     = {http://jmlr.org/papers/v25/19-556.html}
}

\end{document}